\documentclass[10pt,twocolumn,letterpaper]{article}
\usepackage[preprint]{cvpr}
\usepackage{bm}
\usepackage{amssymb}
\usepackage{amsmath}
\usepackage{wrapfig}
\definecolor{cvprblue}{rgb}{0.21,0.49,0.74}
\usepackage[pagebackref,breaklinks,colorlinks,allcolors=cvprblue]{hyperref}
\newcommand\blfootnote[1]{%
\begingroup
\renewcommand\thefootnote{}\footnote{#1}%
\addtocounter{footnote}{-1}%
\endgroup
}

\title{Articulated Kinematics Distillation from Video Diffusion Models \vspace{0.2em}\\
{\normalsize \url{https://research.nvidia.com/labs/dir/akd/}}\vspace{-1.0em}}
\author{Xuan Li$^{1, 2,*}$ \quad Qianli Ma$^{2}$\quad Tsung-Yi Lin$^{2}$ \quad Yongxin Chen$^{2}$ \\
Chenfanfu Jiang$^{1}$ \quad Ming-Yu Liu$^{2}$ \quad Donglai Xiang$^{2}$ \\
$^{1}$ UCLA, $^{2}$ NVIDIA \vspace{-0.2em}
}

\begin{document}

\twocolumn[{%
\renewcommand\twocolumn[1][]{#1}%
\maketitle
\begin{center}
    \centering
    \captionsetup{type=figure}
    \includegraphics[width=0.99\textwidth, trim=0 0 0 0, clip]{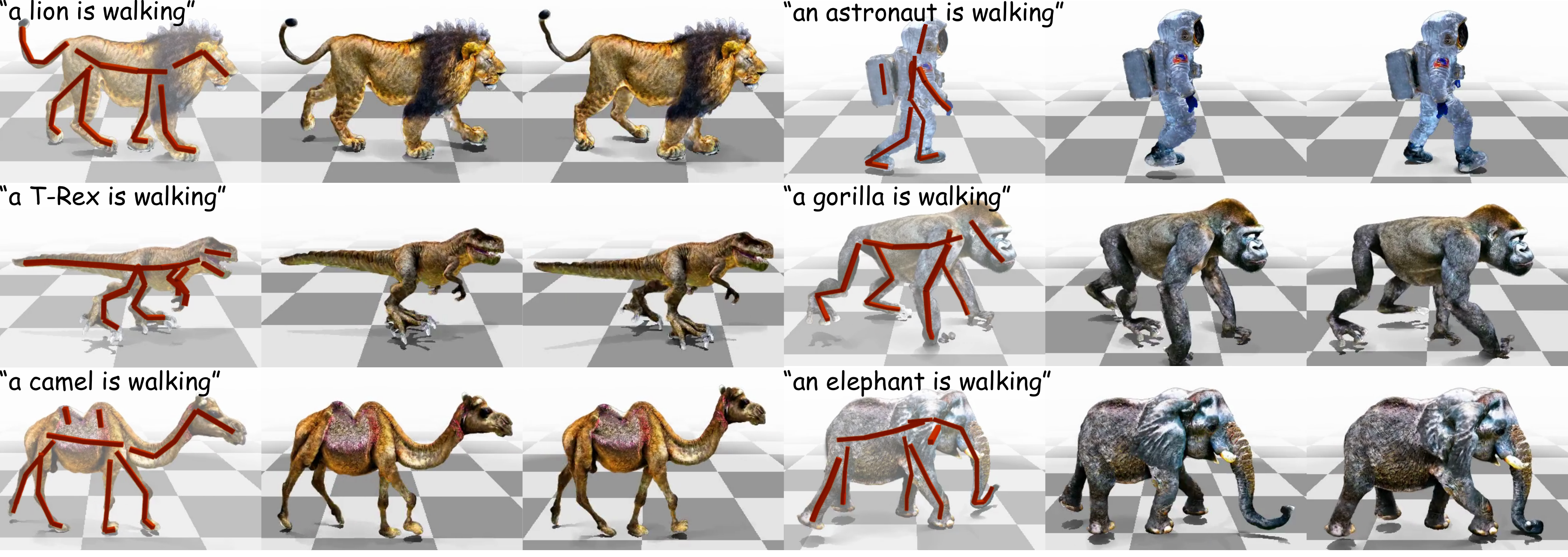}
    \captionof{figure}{By incorporating articulation into static assets, AKD synthesizes realistic motions distilled from large video diffusion models.}
\label{fig:teaser}
\end{center}
}]

\begin{abstract}
\blfootnote{* Work done during an internship at NVIDIA.}
We present Articulated Kinematics Distillation (AKD), a framework for generating high-fidelity character animations by merging the strengths of skeleton-based animation and modern generative models. AKD uses a skeleton-based representation for rigged 3D assets, drastically reducing the Degrees of Freedom (DoFs) by focusing on joint-level control, which allows for efficient, consistent motion synthesis. Through Score Distillation Sampling (SDS) with pre-trained video diffusion models, AKD distills complex, articulated motions while maintaining structural integrity, overcoming challenges faced by 4D neural deformation fields in preserving shape consistency. This approach is naturally compatible with physics-based simulation, ensuring physically plausible interactions. Experiments show that AKD achieves superior 3D consistency and motion quality compared with existing works on text-to-4D generation.
\end{abstract}
\vspace{-1.0em}

\section{Introduction}
\label{sec:intro}

In traditional 3D graphics, a skeleton-based character animation pipeline involves steps like shape modeling, rigging, motion capture, motion retargeting, and editing. As a mature technology, such pipelines can achieve high realism and good controllability over the motion, but they typically require extensive manual work from digital artists, making the process time-consuming and thus hardly scalable. 
Recent advances in video generation models \cite{chen2024videocrafter2, yang2024cogvideox} offer a promising avenue for streamlining the animation authoring process: with a text-to-video model, generating a sequence of character animation only requires a text prompt. 
However, existing video generation models still struggle to generate high-fidelity dynamics for real-world objects because of a lack of 3D information. Common issues include failing to preserve the 3D structure consistency (e.g. number of limbs of a character) during animation, or producing physically implausible articulated motion, such as foot-skating and ground penetration.

Recent works on text-to-4D generation \cite{bahmani20244d, bahmani2025tc4d} leverage these video generation models to distill the learned dynamic motion into consistent 4D sequences. These frameworks commonly rely on neural deformation fields which predict displacements at each location in a pre-defined 3D volume to deform a 3D shape. Animation is thus a temporal sequence of such deformed shapes. While flexible, this approach introduces a large number of Degrees of Freedom (DoFs), making optimization challenging and often resulting in suboptimal quality. Is it possible to have the best of both worlds, where generative models provide extensive knowledge of diverse motions from internet-scale data, while skeleton-based 3D animation allows low-DoF control, permanence of articulated structures, and even physical grounding via simulation?

To answer this question, we introduce Articulated Kinematics Distillation (AKD), a motion synthesis system that bridges traditional character animation pipelines and generative motion synthesis.
Given a rigged 3D asset, we distill articulated motion sequences from a pre-trained video diffusion model using Score Distillation Sampling (SDS). 
The skeleton-based representation simplifies the distillation process by limiting the number of DoFs to that of a few joints, in contrast to all query points in space-time as in the text-to-4D works~\cite{bahmani2025tc4d}. It also offers an effective regularization of the deformation space, enabling the distillation to concentrate on overall motion styles without worrying about maintaining reasonable local structures.
More importantly, the skeleton-based formulation is naturally compatible with physics-based simulation, allowing the generated motion to be grounded by physics-based motion tracking to ensure physical plausibility.

Experiments verify the effectiveness of our design: compared to previous text-to-4D methods, our framework produces results with better 3D shape consistency and more expressive motions. 
We summarize our contributions:
\begin{itemize}
    \item We introduce a novel text-driven motion synthesis framework for static 3D assets, combining articulated rigging systems and large video diffusion models.
    \item We demonstrate that incorporating non-uniform ground renderings enhances the video model’s adherence to basic physics between the character and the ground.
    \item Extensive experiments show that our generated motions exhibit higher quality than the state-of-art methods that can synthesize long-trajectory motions.
    \item Our generated motion can be used in physics-based motion tracking with differentiable physics to further boost its physical realism.
\end{itemize}

\section{Related Work}
\label{sec:related}
\paragraph{Deformable Gaussian Splatting}

In recent years, different types of 3D representations have been introduced to facilitate the reconstruction and generation of 3D/4D scenes, such as neural fields \cite{mildenhall2021nerf}, iNGP \cite{muller2022instant}, and 3D Gaussian Splatting \cite{kerbl20233d}. Among these representations, 3D Gaussian Splatting is particularly suitable for representing dynamic scenes due to its explicit nature \cite{xie2024physgaussian} compared to the NeRF based on neural implicit fields \cite{park2021nerfies, qiao2023dynamic}, whose deformations are achieved by bending rendering rays \cite{pumarola2021d, feng2024pie}. This advantage of 3DGS has sparked a lot of works on 4D reconstruction and modeling from multi-view input, including general scenes \cite{yu2024mip}, facial avatars \cite{xu2024gaussian,giebenhain2024npga}, and full-body avatars \cite{shao2024splattingavatar, moreau2024human, kocabas2024hugs, hu2024gauhuman, liu2024gva, zheng2024physavatar}, where 3D Gaussian kernels are bound to an articulated human model SMPL \cite{loper2023smpl} through learned skinning weight. We adopt GS as our 3D shape representation, which naturally allows SDS gradients to smoothly propagate through the articulated deformation and rendering pipeline.

\begin{figure*}
    \centering
    \includegraphics[width=\linewidth]{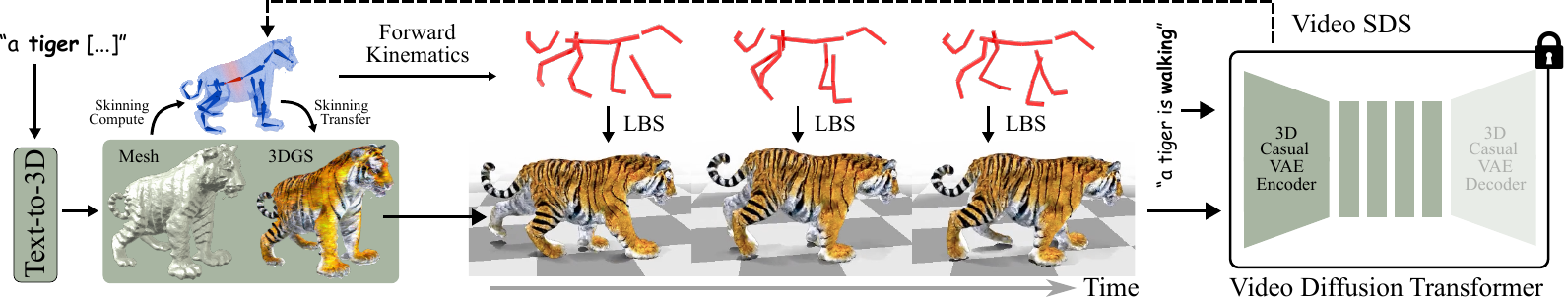}
    \caption{\textbf{Pipeline.} We novelly incorporate articulated skeletons into generative motion synthesis. With the low-dimensional parameterization of motions (a sequence of joint angles for articulated bones), the synthesis can focus more on motion modes instead of local-scale deformations. Given a text prompt, we use a text-to-3D method to generate a 3D asset. The asset is deformed by the skeleton and differentiably rendered into videos. The SDS gradient is evaluated by a pre-trained video diffusion transformer and backpropagated to joint angles.}
    \label{fig:pipeline}
\end{figure*}

\vspace{-1em}
\paragraph{Articulated Motion Reconstruction}
A closely related topic to our work is the reconstruction of articulated motions of deformable objects in under-constrained settings, especially from a monocular view. As a special case, the reconstruction of human body poses \cite{kanazawa2018end,xiang2019monocular,goel2023humans} benefits strongly from the availability of dedicated deformable models such as SMPL \cite{loper2023smpl} and the abundance of 2D/3D pose data \cite{lin2014microsoft,mahmood2019amass}. In contrast, the reconstruction of general objects, such as animals and humans in loose clothing, has remained more challenging due to the inherent 3D ambiguity from a monocular view and a lack of reliable priors for their articulation. One line of works following BANMO \cite{yang2022banmo,yang2023reconstructing,yang2023ppr,lei2024gart,tan2024dressrecon} solve for a static 3D template and articulated motion simultaneously from a monocular video by leveraging various image measurements such as segmentation, optical flow, and DensePose. Several works from another line \cite{yao2022lassie,yao2023hi,wu2023magicpony,li2024learning,jakab2024farm3d} train a neural network that predicts the shape template of animals and their body deformation conditioned on a single input image in a weakly-supervised manner. Ponymation \cite{sun2025ponymation} learns a motion VAE for horse motions from a collection of horse videos. There are also works \cite{tevet2023human, raab2024single, yuan2023physdiff, jiang2023motiongpt} that directly train generative models on articulated motions. Instead of extracting articulation from a particular input, our method focuses on generating novel skeleton motion utilizing video diffusion priors that have learned general visual knowledge.

\vspace{-1em}
\paragraph{Generating 4D and Physics-Based Dynamics}
Generating 4D content is inherently challenging as it demands high consistency not only along the temporal axis to maintain motion, but also across multiple viewpoints to ensure spatial and structural accuracy in the generated content. Some works attempt to directly build priors models in the 4D domain, including diffusion models \cite{xie2024sv4d, jiang2024animate3d} and reconstruction model \cite{ren2024l4gm}, where the limited availability of 4D data pose a challenge.
Therefore, a large body of works \cite{singer2023text, ren2023dreamgaussian4d, ling2024align, bahmani20244d, jiang2023consistent4d, zheng2024unified, zeng2025stag4d, uzolas2024motiondreamer} distill 4D motion from a combination of generative models that operate in lower dimensions, including images, videos, and multi-view images (3D), which is a difficult problem to due to spasity of supervision, noisy gradient from SDS, and high-degree of freedom in the optimization variables. Some works aim to improve the controllability of the 4D generation by introducing conditioning of trajectory \cite{wu2025sc4d} or sparse control points \cite{wu2025sc4d}. None of the works above allow the generated contents to be grounded in a physics-based manner. The exceptions are PhysGaussian \cite{xie2024physgaussian} and PhysDreamer \cite{zhang2025physdreamer}, which models scene deformation by a Material-Point-Method (MPM) simulator. Such a formulation is more suitable for modeling the volumetric solid deformation and fluid dynamics, whereas our work focuses on articulated motion which can be simulated more efficiently by a skeleton-based representation.

\section{Background}
\subsection{Deformable Gaussian Splatting}
3D Gaussian Splatting (3DGS) \cite{kerbl20233d} utilizes a set of 3D Gaussian kernels to represent a 3D scene. These kernels can be rendered into images using a customized differentiable rasterizer. Specifically, each kernel is defined by a set of parameters, ${\sigma_p, \bm x_p, \bm \Sigma_p, \mathcal{C}_p}$, where $\sigma_p$ denotes the opacity, $\bm x_p$ and $\bm \Sigma_p$ represent the center and covariance matrix of the Gaussian ellipsoid, and $\mathcal{C}_p$ is the coefficient set of spherical harmonics that determines the view-dependent colors of the kernel. Given a camera view, the color of a pixel is rendered by $\alpha$-blending all kernels along the ray direction:
\begin{equation} \bm C = \sum_i \bm c_i \alpha_i \prod_{j=1}^{i-1} (1-\alpha_j), \end{equation}
where $i$ is the sorted index of the kernels in ascending order of $z$-depth, $\bm c_i$ represents the kernel's color under the given view, evaluated from spherical harmonics, and $\alpha_i$ is the opacity $\sigma_i$ weighted by the kernel's 3D Gaussian distribution. Given a set of images with known camera parameters, we can reconstruct the 3D scene using only RGB loss.

Following PhysGaussian \cite{xie2024physgaussian}, given a warping function $\bm \phi: \mathbb R^3 \rightarrow \mathbb R^3$ (defined in Sec.~\ref{sec:method-rig}), we can deform the shape of each Gaussian kernel using the local linearization of $\bm \phi$ at its center $\bm x_p$: $\bm \phi(\bm x) \approx \bm \phi(\bm x_p) + \nabla \phi(\bm x_p)(\bm x - \bm x_p)$, resulting in:
\begin{equation} \label{eq:deformation}
    \tilde{\bm x}_p = \phi(\bm x_p), \quad \tilde{\bm \Sigma}_p =  \nabla \phi(\bm x_p) \bm \Sigma_p \nabla \phi(\bm x_p)^T,
\end{equation}
where $(\bm x_p, \bm \Sigma_p)$ and $(\tilde{\bm x}_p, \tilde{\bm \Sigma}_p)$ denote the Gaussian parameters before and after the deformation respectively.
For simplicity, we can hold opacities and spherical harmonics constant. Our deformation system adheres to this convention to generate Gaussian kernels in motion.

\subsection{Video Score Distillation Sampling}
The Score Distillation Sampling (SDS) method, proposed to distill 3D models from text-to-image priors \cite{poole2023dreamfusion}, uses an approximated gradient of the diffusion loss \cite{ho2020denoising}:
\begin{equation}
    \mathcal{L}_{\text{Diff}}(\theta, \bm z,  y) = \mathbb{E}_{t, \bm\epsilon} \left[ w(t) \left\| \hat{\bm\epsilon}  - \bm\epsilon \right\|_2^2 \right],
\end{equation}
where $t\sim \mathcal{U}(0,1)$ is the diffusion time step, $\bm\epsilon \sim \mathcal{N}(0, \mathbf{I})$ is a Gaussian noise, 
$\bm z_t = \sqrt{\alpha_t} \bm z + \sqrt{1 - \alpha_t} \bm\epsilon$ is the noisy image at step $t$, $\bm z$ is the rendered image, $y$ is the input text, and $\hat{\bm\epsilon} = \bm\epsilon_{\theta}(\bm z_t; t, y)$ is the noise predicted by the diffusion model. Omitting the gradient through the noise-predicting module, the gradient of $\mathcal{L}$ w.r.t. $\bm z$ is given by:
\begin{equation}
    \nabla_{\bm z}  \mathcal{L}_{\text{SDS}}(\bm z,  y) = \mathbb{E}_{t, \bm\epsilon} \left[ w(t)  \left(\hat{\bm\epsilon} - \bm\epsilon \right)\right].
\end{equation}
This pixel-level gradient is then backpropagated to the 3D model parameters via a differentiable renderer, guiding the generation of a 3D representation consistent with the input text. Videos, which are essentially 3D image tensors with an additional temporal axis, are handled similarly; the SDS gradient can be backpropagated to 4D representations.

\section{Method}

We present Articulated Kinematics Distillation, a text-driven articulated motion synthesis system for animating 3D assets, as illustrated in \cref{fig:pipeline}.
We assume a rigged 3D asset as input. For results in this paper, we use an off-the-shelf text-to-3D generator, although assets from photogrammetry or created by artists should work similarly. We then convert the asset into a dual mesh-3DGS representation: the mesh is used to compute appropriate skinning weights, and the 3DGS supports differentiable deformations and rendering during SDS optimization. We rig the asset by manually embedding an articulated skeleton, and compute Linear Blend Skinning (LBS) weights, which we transfer to nearby Gaussian kernels. To generate the motion, we optimize a sequence of joint angles within the articulation tree, which pass through a differentiable Forward Kinematics (FK) and 3DGS rasterization pipeline to render a video sequence. The rendered video is then fed into a pre-trained video diffusion model, which provides guidance for adjusting joint angles at each frame to produce text-aligned motion. In the following sections, we present the technical details of each component.

\subsection{Rigging and Skinning for Gaussian Splatting}
\label{sec:method-rig}
The rigging system is widely used in the animation industry, where 3D character meshes are controlled via an embedded articulated skeleton. Artists manipulate joint angles between connected bones to create 3D animations. The transformations of the deformed bones are evaluated using Forward Kinematics \cite{denavit1955kinematic} of the skeleton tree, and then mesh vertices are moved by interpolating transformations from nearby bones. The most common interpolation scheme is Linear Blend Skinning (LBS) \cite{magnenat1988joint}, which defines the deformation function as:
\begin{equation}
    \bm \phi(\bm x) = \sum_{i=1}^B w_i (\bm R_i \bm x + \bm T_i),
\end{equation}
where $B$ is the number of bones, $(\bm R_i, \bm T_i)$ is the rigid transformation of bone $i$, and ${w_i}$ is the skinning weight, satisfying $\sum_i w_i = 1$.
This deformation field is usually applied to mesh vertices, but it is defined on the entire 3D space, meaning that it can also deform Gaussian kernels following \cref{eq:deformation}, where $\nabla \phi(\bm x) = \sum_{i=1}^B w_i \bm R_i$.
We transfer skinning weights to Gaussian kernels by barycentric interpolation of skinning weights at their nearest points on the template mesh. For the assets generated by text-to-3D methods, we use Pinocchio \cite{baran2007automatic} to compute skinning weights for mesh vertices automatically. Please refer to our supplementary document for more details.

\subsection{Rendering} \label{sec:rendering}
For each deformed frame of the 3DGS asset, we use the differentiable Gaussian Splatting rasterizer \cite{kerbl20233d} to render it into an image. These rendered frames are then concatenated sequentially to form a video tensor.

\vspace{-1em}
\paragraph{Ground Rendering.} Unlike previous text-to-4D frameworks \cite{zheng2024unified,bahmani2025tc4d} which often render environments in nearly uniform colors, we incorporate a checkerboard ground. We find that this setup can provide the distillation with clues about the interaction between assets and the ground. It helps reduce relative motions between contacting parts and the ground, and helps keep assets grounded. We render the checkerboard ground as a background layer by ray casting and blend it with the rendering of the asset. For each ray corresponding to a pixel, the color is determined by the intersection point between the ray and the ground; if there is no intersection, the color is set to a pre-assigned value. 
We also set the opacities of kernels located below the ground to zero to account for occlusion. 

\vspace{-1em}
\paragraph{Camera Trajectory.} We let the camera smoothly follow the bounding box center of the deformed shape in each frame, as the video model we use often generates object-centric videos. It is equivalent to moving the entire scene in the opposite direction while keeping the camera view fixed. 

\subsection{SDS for V-Prediction Diffusion Models}
The video model we use, CogVideoX-5B \cite{yang2024cogvideox}, is a Diffusion Transformer (DiT) trained with a $v$-prediction formulation \cite{salimans2022progressive}. The diffusion loss is defined as:
\begin{equation}
    \mathcal{L}_{\text{Diff}}(\theta, \bm z,  y) = \mathbb{E}_{t, \bm\epsilon} \left[ w(t) \left\| \bm z - \hat{\bm z} \right\|_2^2 \right],
\end{equation}
where $\hat{\bm z} = \sqrt{\alpha_t} \bm z_t - \bm v_{\theta}(\bm z_t; t, y)$ is the reconstruction based on the predicted velocity by the diffusion model $\bm v_{\theta}$. Omitting the gradient through the velocity-predicting transformer, the SDS gradient is evaluated as:
\begin{equation}
     \nabla_{\bm z}  \mathcal{L}_{\text{SDS}}(\bm z,  y)  = \mathbb{E}_{t, \bm\epsilon} \left[ w(t) \left( \bm z - \hat{\bm z} \right) \right].
\end{equation}
Derivations are provided in the supplementary document.

Modern video diffusion models are trained in latent space by encoding raw videos using Variational Autoencoders (VAE). Correspondingly, the SDS gradient is computed on the latent codes, and then needs to be back-propagated through the VAE encoder into the pixel space. This process can be extremely memory-intensive, especially when handling a large number of frames. To reduce the memory footprint, we employ the gradient checkpointing techniques \cite{chen2016training}. These techniques save memory by selectively storing intermediate activations and recomputing them during the backward pass, making it feasible to perform SDS optimization with large DiTs.

\subsection{Optimization}
Given all the components, we now present the optimization process of our framework. Our approach uses 3-DoF compound spherical joints to connect bones, where each DoF represents the rotation angles around one of three linearly independent axes. The optimizable variables for each asset consist of $\Theta = \left\{\{\mathcal{A}_i^j\}_{j=1}^{B-1}, \mathcal{T}_i\right\}_{i=0}^{F-1}$, where $F$ is the number of frames, $\mathcal{A}_i^j$ is the 3D angle vector for joint $j$ at frame $i$, and $\mathcal{T}_i$ denotes the 6-DoF rigid transformation of the root bone in the articulation tree at frame $i$. We use the following loss function for SDS optimization:
\begin{equation}
    \mathcal{L} = \mathcal{L}_{\text{SDS}} + \lambda_1 \mathcal{L}_{\text{smooth}} + \lambda_2 \mathcal{L}_{\text{ground}}.
\end{equation}
Besides SDS loss, we incorporate two regularizers: a smoothness penalty $\mathcal{L}_{\text{smooth}}$ over time to encourage time-consistent deformations, and a ground penetration penalty $\mathcal{L}_{\text{ground}}$ to enforce the assets to stay above the ground. 

Since our parameters have physical meanings, we can directly enforce smoothness on the control parameters, which is defined as the mean absolute error (MAE) of the Laplacian of the control parameters w.r.t. time, excluding the first and last frames:
\begin{equation}
\small 
\mathcal{L}_{\text{smooth}} = \operatorname{MAE} (\Delta_t \Theta),\ \   (\Delta_t \Theta)_{i} = \Theta_{i-1} - 2 \Theta_i + \Theta_{i+1}.
\end{equation}
This loss helps ensure that changes in the parameters across consecutive frames are gradual, annealing the noisy nature of the SDS loss. 
The ground penetration loss is defined as 
\begin{equation}
   \mathcal{L}_{\text{ground}} = \frac1{|V|}\sum_{\bm v \in V}\operatorname{max}(-\bm v^y,0).
\end{equation}
Here, we conceptualize bones as a set of transformed cuboids, and $V$ represents the set of vertices of these cuboids. This loss penalizes any penetration of the asset below the ground plane by applying a penalty proportional to the depth below ground for each vertex.

We focus on motion synthesis for legged characters, including animals and humanoids, moving across the ground. In our experiments, we initialize the asset’s forward motion along $x$-axis by setting an initial displacement as $\mathcal{T}_i^x = vi$ for some constant velocity $v$. The pace and trajectories of the assets can be further optimized during the distillation.

\subsection{Physics-Based Motion Tracking}
The articulated skeletons embedded in the asset can be deployed in articulated rigid body simulators. To ground the synthesized motion in physics, we can further project the distilled motion trajectory to the nearest solution achievable in physics-based tracking in a simulation environment. We accomplish this generation-to-simulation transition by searching for a physical joint control sequence that minimizes the difference between simulated and synthesized bone trajectory.

Specifically, we treat each bone as a rigid cuboid and use compound joints to connect bones, which is consistent with the motion DoFs during motion distillation. We use a semi-implicit articulated rigid body simulator to simulate the skeleton under gravity and ground collision. The simulation process can be abstracted as a state transition $\mathcal{S}$:
\begin{equation}
    [\bm q^{k+1}, \dot{\bm q}^{k+1}] = \mathcal{S}([\bm q^{k}, \dot{\bm q}^{k}], \Delta t, \bm \tau^k)
\end{equation}
where $\bm q^{k}$ is the concatenation of generalized coordinates of bones at the time step $k$, which represent 6-DoF rigid transformations, $\dot{\bm q}^{k}$ is the generalized velocity, $\Delta t$ is the simulation time step size, and $\bm \tau^k$ is active joint torques applied to the connected bones at joints. We use a Proportional-Derivative (PD) controller to provide active joint torques, where the torque at joint $j$ around axis $l$ is given by:
\begin{equation}
    \bm \tau^k|_{l} = k_e (\hat{\theta}_{jl} - \theta_{jl}) - k_d \dot{\theta}_{jl}
\end{equation}
where $\hat{\theta}_{jl}$ is the control variable, and $\theta_{jl}$ is the current joint angle. This formulation tends to pull the joint angle $\theta_{jl}$ towards $\hat{\theta}_{jl}$ subject to damping. $k_e, k_d$ are globally defined parameters that control joint elasticity and damping, respectively. The whole simulation process can be implemented using AutoDiff frameworks such as Warp \cite{warp2022}. Altogether, we use the following motion-tracking loss:
\begin{equation}
\footnotesize
   \mathcal{L}_{\text{track}} (\hat{\Theta}, \dot{\bm q}^0) = \frac{1}{F-1}\sum_{i=1}^{F-1}\|\bm q^{iN} - \hat{\bm q}^{i}\|_1 + \lambda_3 \operatorname{MAE}(\Delta_t \hat{\Theta}).
\end{equation}
Here, $N$ is the ratio between the frame time of the target motion $\{\hat{\bm q}^{i}\}$ and the simulation time step $\Delta t$, and $\hat{\Theta}$ is the concatenation of control variable per simulation step. Like in the distillation, we add a smoothness regularizer. We also optimize the initial generalized velocity $\dot{\bm q}^0$ as well. The simulation starts from $\bm q^0 = \operatorname{Proj}(\hat{\bm q}^0)$, where $\hat{\bm q}^0$ is vertically shifted such that the lowest bone touches the ground. 

However, $N$ is typically in the hundreds for explicit simulators, resulting in thousands of simulation steps in total. This can cause severe gradient explosion issues during backpropagation. A common technique to alleviate this issue is to apply gradient clipping before each optimizer step. However, the explosion may occur multiple times throughout backpropagation, which can make the final gradient useless. To address this, we employ a fine-grained gradient clipping strategy, applying gradient clipping every few tens of backward steps. We achieve this by wrapping chunks of substeps into a PyTorch layer and manually defining the backpropagation logic.

\begin{figure*}[t!]
    \centering
    \includegraphics[width=\linewidth]{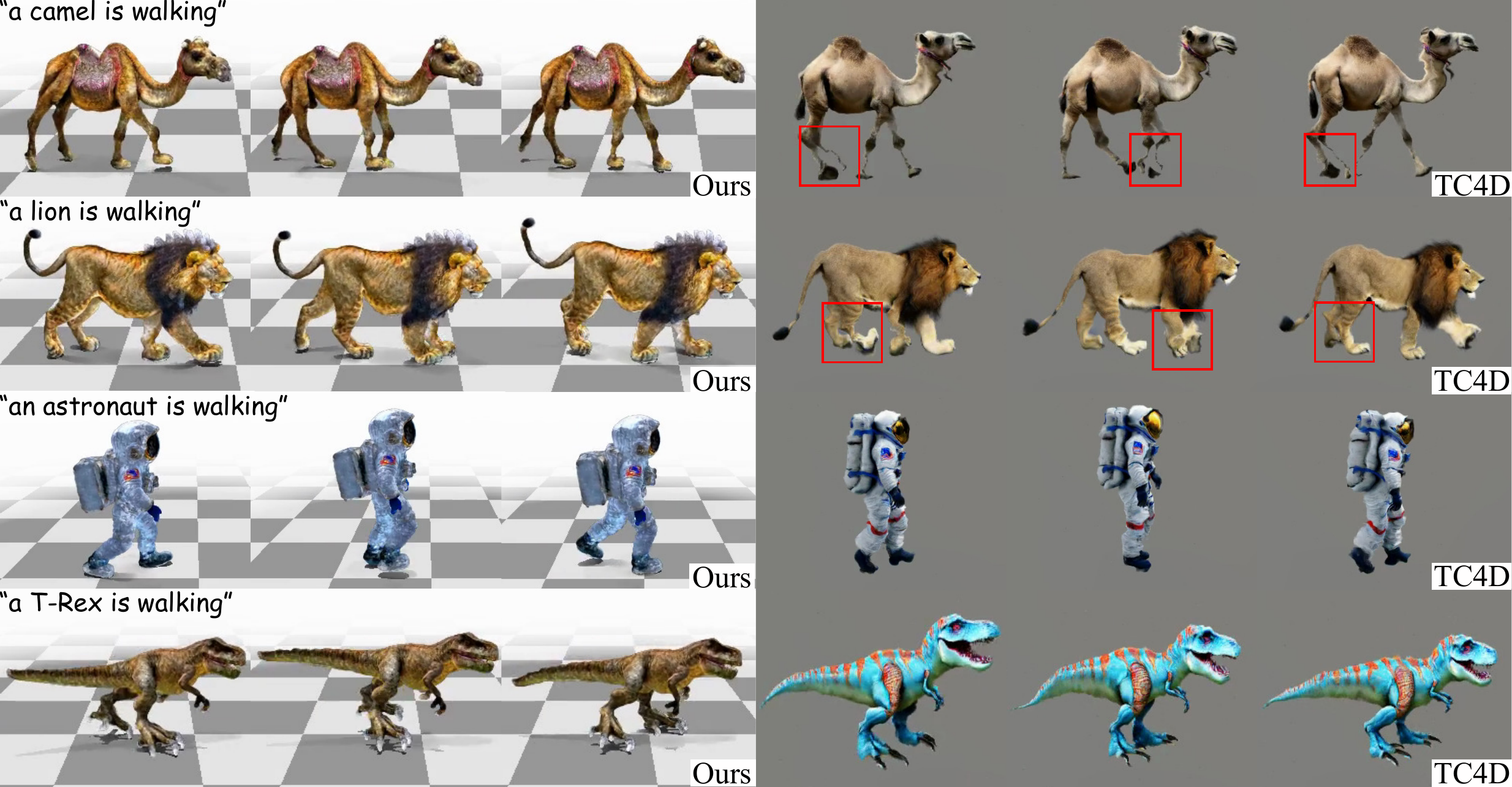}
    \caption{Qualitative comparisons with TC4D. The blurry artifacts generated by TC4D are highlighted. TC4D often fails to produce alternating leg movements (e.g., in the astronaut example), or shows limited local-scale motion (e.g., in the T-Rex example).
    }
    \label{fig:comparison}
\end{figure*}

\section{Implementation}

We use an off-the-shelf SDS-based text-to-3D system, Tet-Splatting \cite{gu2024tetrahedron}, to generate static 3D mesh assets from texts. The text-to-3D module is not the focus of this work and can be replaced by any suitable framework. We set up a skeleton inside the mesh manually in Blender \cite{blender} using its armature system. This manual setup typically takes only a few minutes per instance. We convert the generated mesh into the Gaussian Splatting representation by applying 3DGS reconstruction \cite{kerbl20233d} to images of the mesh from 200 random views, starting from mesh vertices as initial kernel centers.

We adopt threestudio \cite{threestudio2023} for SDS optimization, and CogVideoX-5B \cite{yang2024cogvideox} as the video diffusion model. By leveraging the gradient checkpointing mechanism of the VAE, all our experiments can be run on a single NVIDIA A100-40GB graphics card. We adopt forward kinematics and articulated rigid body simulation from Warp \cite{warp2022} and integrate it with PyTorch for data and gradient interchange. Each asset undergoes 10,000 iterations of SDS optimization, requiring approximately 25 hours.

\begin{figure*}[t!]
    \centering
    \includegraphics[width=\linewidth]{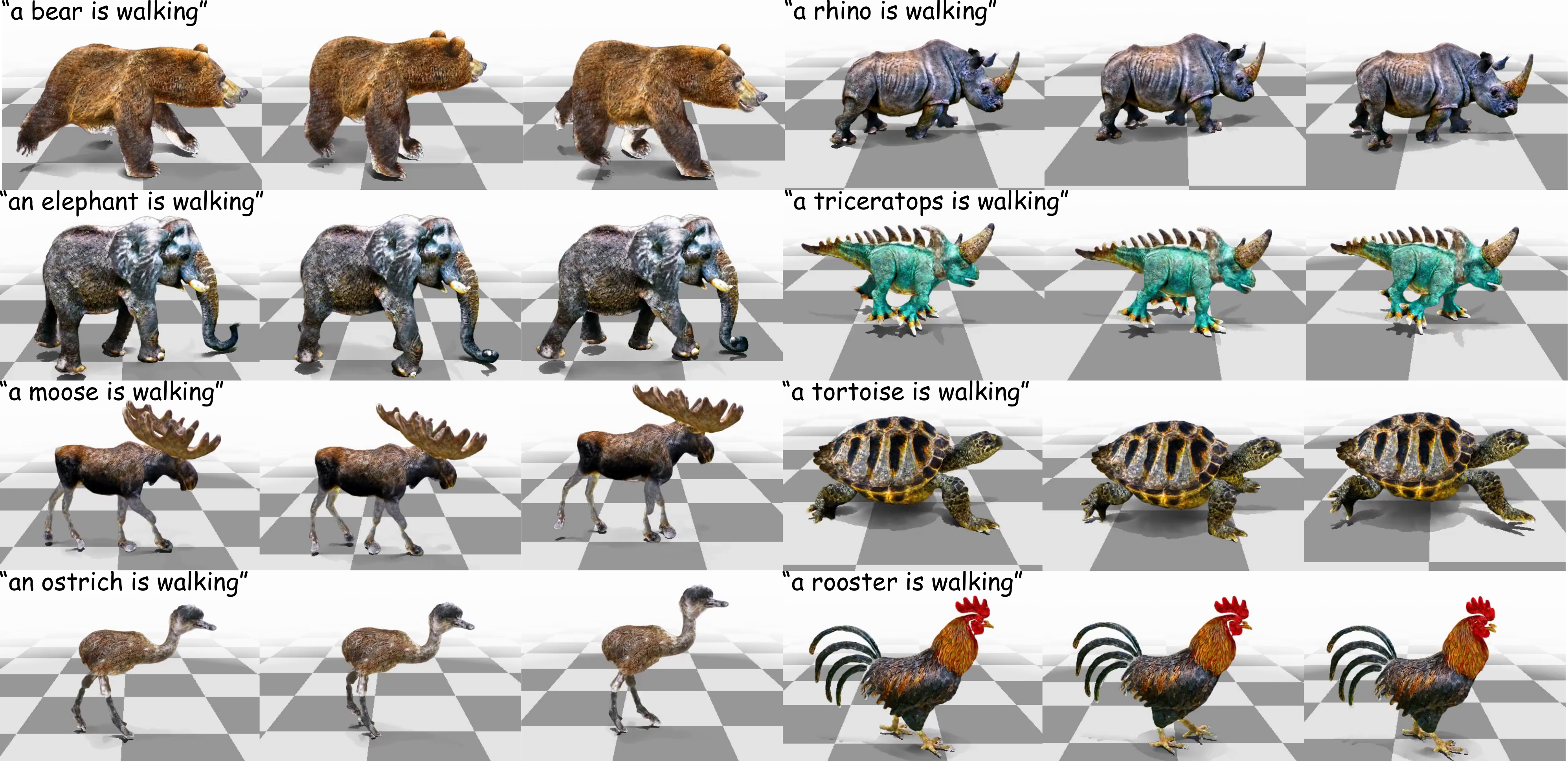}
    \caption{Examples of our synthesized motions. \vspace{-0.5em}}
    \label{fig:more_result}
\end{figure*}

\section{Experiment}
\subsection{Quantitative Comparisons}
We generate 29 static assets, including animals and humanoids, and compare our method with the state-of-the-art open-sourced approach, TC4D \cite{bahmani2025tc4d}, that can synthesize long-trajectory motions. TC4D moves the bounding box of the object along a predefined path, while our method only initializes a path and allows adjustments based on the guidance from the video model. Other text-to-4D frameworks, such as those in \cite{bahmani20244d, zheng2024unified}, are limited to synthesizing local-scale motions, which tend to have constrained magnitudes. We compare our method (without physics-based tracking for fairness) with TC4D using both automated video scores and a user study.
\subsubsection{Metrics}
\paragraph{Automated Metric} We utilize VideoPhy \cite{bansal2024videophy} to automatically score videos. This model scores two types of metrics: Semantic Adherence (SA), which assesses the alignment between the text and the video content, and Physical Commonsense (PC), which evaluates whether the video adheres to basic physical laws. We observed that the evaluation of PC scores tends to concentrate on the local deformations of objects. For a fair comparison with TC4D using this metric, we exclude ground rendering when using this metric.
\vspace{-1.0em}
\paragraph{User Study}
We recruited 20 human evaluators to compare video clip pairs generated by AKD and TC4D using the same text prompts. We asked the evaluators to assess each comparison across several aspects: Motion Amount (MA), Physical Plausibility (PP), and Text Alignment (TA). Evaluators were instructed to choose the clip they felt performed better for each aspect.

\subsubsection{Result}
\begin{wraptable}{r}{0.2\textwidth}
\vspace{-0.5em}
\hspace{-2em}
\resizebox{0.23\textwidth}{!}{
\begin{tabular}{lll}
\toprule
     & \multicolumn{1}{c}{SA} & \multicolumn{1}{c}{PC} \\ \toprule
TC4D & $0.40 \pm 0.34$        & $0.31 \pm 0.15$        \\ \hline
Ours & $\bm{0.81 \pm 0.26}$        & $\bm{0.39 \pm 0.17}$        \\ \bottomrule       
\end{tabular}
}
\vspace{-0.3em}
\end{wraptable}
The automatic scores evaluated by VideoPhy are provided in the inset table where the means and standard deviations of each score are reported. Our method outperforms TC4D in terms of both metrics. The user study results show that our method was preferred in $51\%$, $53\%$, and $53\%$ of the comparisons in terms of MA, PP, and TA, respectively. These results indicate that our method can synthesize motions in better quality. 

\subsection{Qualitative Comparisons}
In \cref{fig:comparison}, we visualize several comparisons between TC4D and ours. We refer readers to the supplemental video for complete results. Note that appearance is not our focus because the 3D assets can be generated by any text-to-3D method. We observe that TC4D rarely produces alternating leg movements, and areas where the legs converge often appear blurry. In contrast, our method, with its low-DoF parameterization, effectively maintains the geometric structure of each part within the assets during deformations.

\subsection{Physics-based Motion Tracking}
\begin{figure}[t!]
    \centering
    \includegraphics[width=\linewidth]{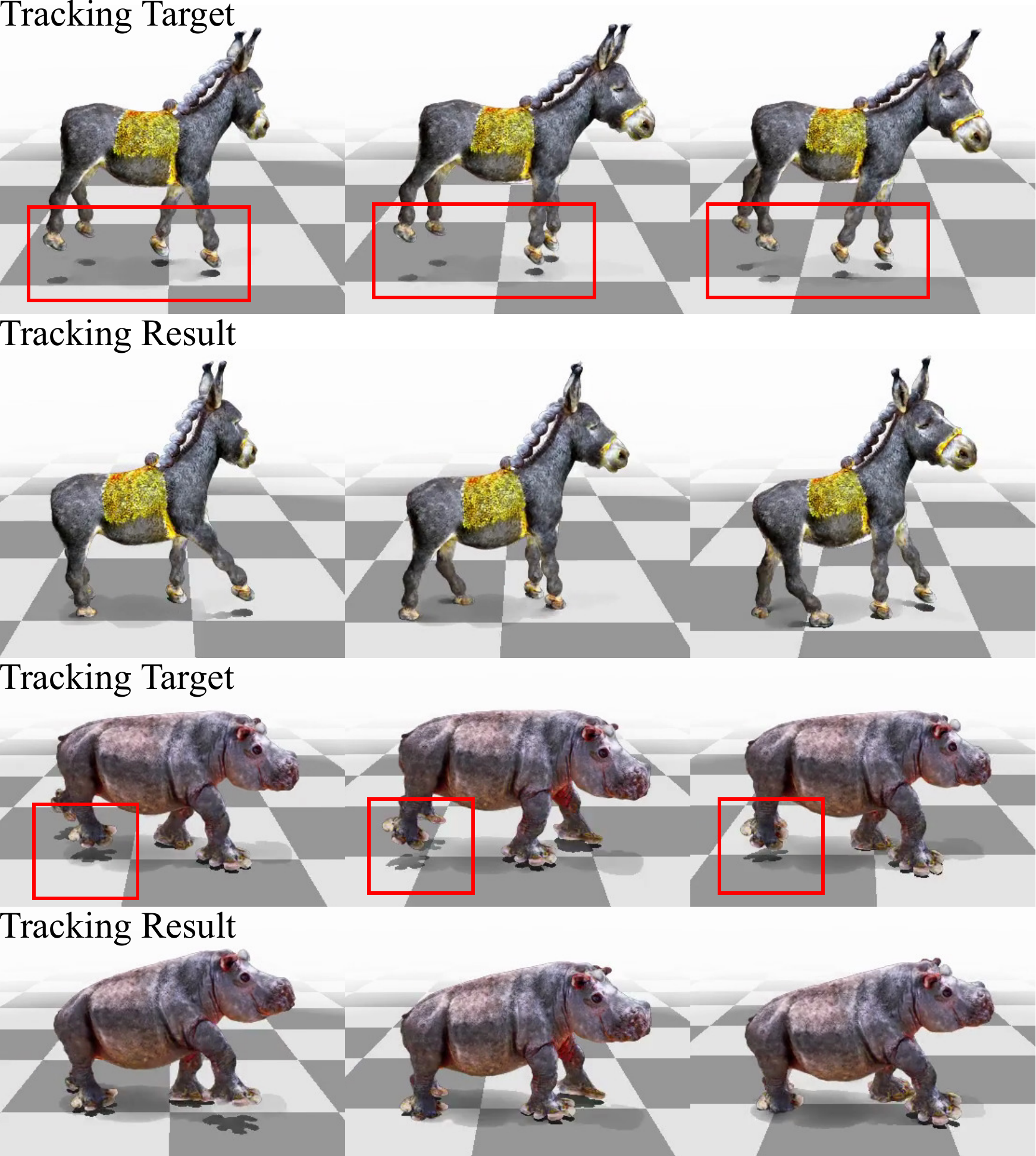}
    \caption{We use physics-based motion tracking to project synthesized motions onto physics-grounded trajectories. \vspace{-2.0em}}
    \label{fig:motion-tracking}
\end{figure}

The synthesized articulated motions may still float above the ground or slide relative to it, as video diffusion models cannot provide guidance for such fine-grained placements. However, the overall styles of motions are well captured. In \cref{fig:motion-tracking}, we present examples where the synthesized motions appear above the ground, while the physics-based tracking results adhere to gravity and have frictional contact with the ground, yet still track the target motion styles. For more results, we refer the reader to the supplemental video.

\subsection{Synthesis Diversity}
\paragraph{Asset Diversity.}
In \cref{fig:more_result}, we visualize more examples of motions by our method on different assets with varying articulated topologies. We refer readers to the supplemental video for animations.

\begin{figure}
    \centering
    \includegraphics[width=\linewidth]{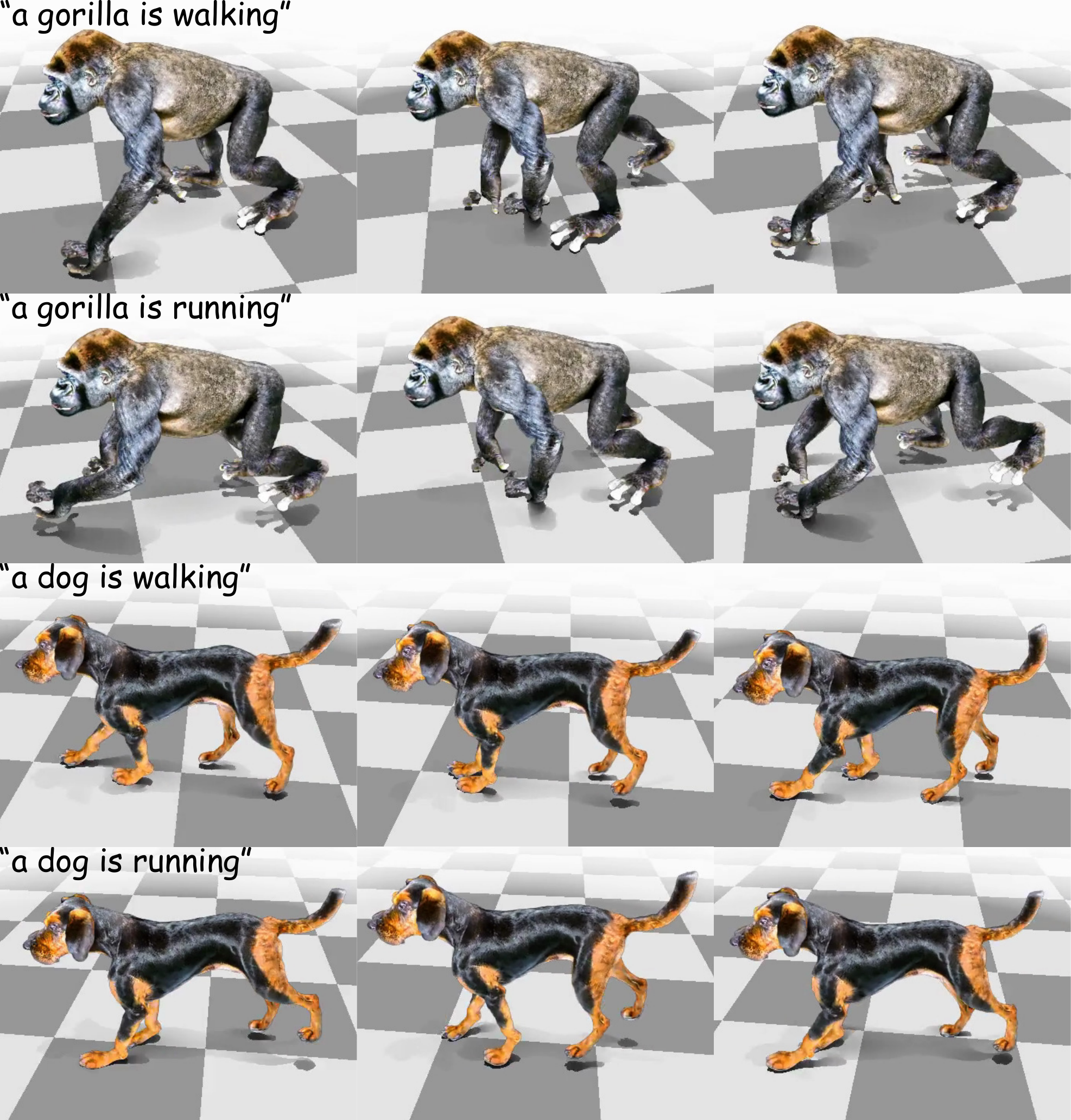}
    \caption{Our method supports synthesizing different motions based on varying text descriptions.}
    \label{fig:diversity}
\end{figure}

\vspace{-1.0em}
\paragraph{Motion Diversity.}
Our method supports the synthesis of different motion modes based on different text prompts. In \cref{fig:diversity}, we illustrate the differences between ``walking" and ``running" behaviors. For quadrupeds, the limbs on each side alternate during walking, whereas they move in sync during running.

\subsection{Ablation Studies}
\begin{figure}[t!]
    \centering
    \includegraphics[width=\linewidth]{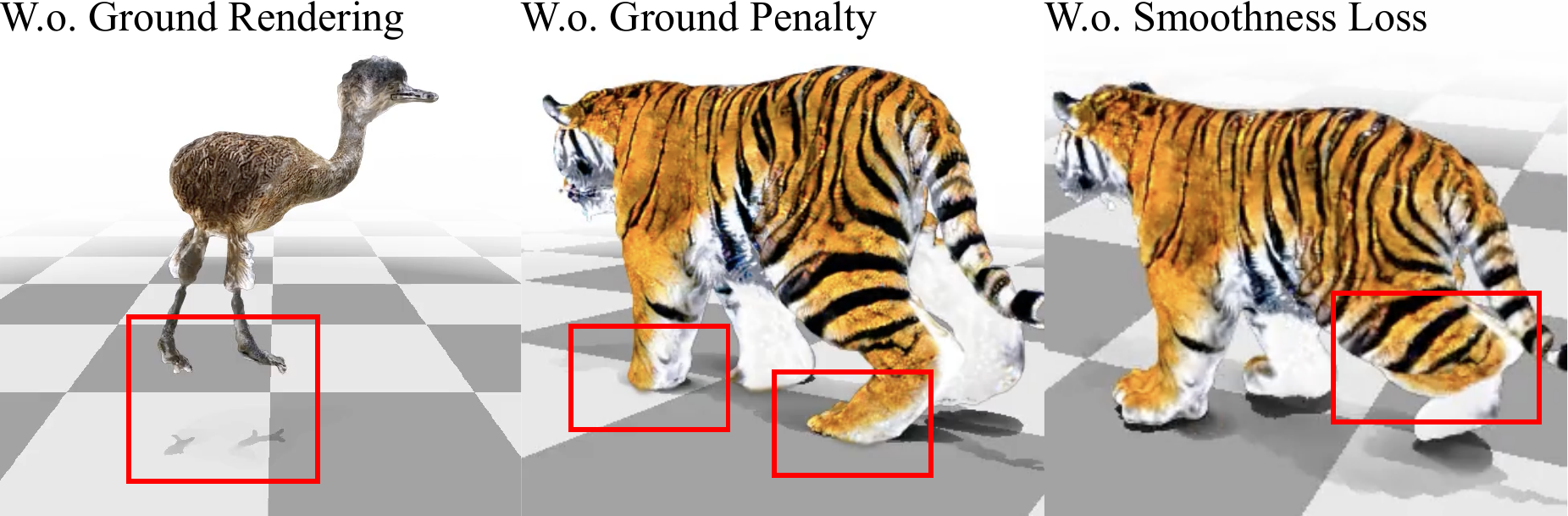}
    \caption{Ablation studies on ground rendering, ground penalty loss, and the smoothness loss. The artifacts are highlighted.\vspace{-1em}}
    \label{fig:ground}
\end{figure}

\paragraph{Components in Training.}
\begin{wraptable}{r}{0.22\textwidth}
\vspace{0.3em}
\hspace{-2em}
\resizebox{0.25\textwidth}{!}{
\setlength{\tabcolsep}{3pt}
 \begin{tabular}{cccccc}
 \hline
      & Ostrich & \begin{tabular}[c]{@{}c@{}}w.o.\\ground \end{tabular} & Tiger & \begin{tabular}[c]{@{}c@{}}w.o.\\$\mathcal{L}_{\text{ground}}$ \end{tabular}  & \begin{tabular}[c]{@{}c@{}}w.o.\\$\mathcal{L}_{\text{smooth}}$ \end{tabular}   \\
      \hline
     SA & \textbf{0.982} & 0.777 & \textbf{0.989} & 0.979 & 0.984\\
     PC & \textbf{0.294} & 0.107 & \textbf{0.321} & 0.269 & 0.294\\
     \hline
 \end{tabular}
 }
 \vspace{-0.7em}
\end{wraptable}
We conducted ablation studies on several components in addition to SDS loss: ground rendering, ground penalty loss, and smoothness loss, each targeting specific types of artifacts. The visualizations are presented in \cref{fig:ground}. The automated quantitative metrics evaluated using VideoPhy are reported in the inset table. Without ground rendering, the synthesized motion appears above the ground because the video model lacks information about the ground's location. The absence of the ground penalty loss results in penetration into the ground. The smoothness loss can encourage time consistency across frames; without it, dramatic and unnatural deformations may occur in certain frames. Both SA and PC scores decrease when one of these component is removed.

\begin{figure}
    \centering
    \includegraphics[width=\linewidth]{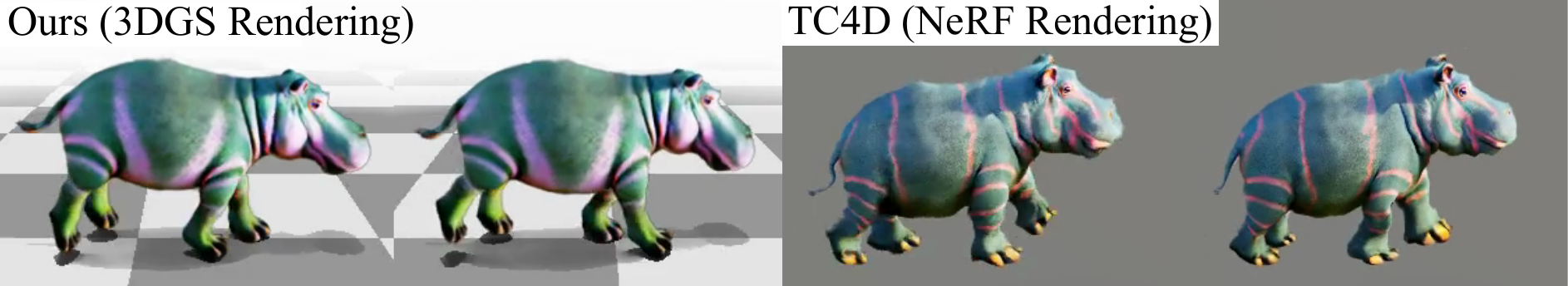}
    \caption{Ablation on the text-to-3D module. We extract an asset from TC4D and achieve a comparable appearance. \vspace{-1em}}
    \label{fig:text23d}
\end{figure}

\vspace{-1em}
\paragraph{Text-to-3D Module.} In this paper, the text-to-3D module is not our focus. We use Tet-Splatting because of its efficiency. This module can be replaced by arbitrary text-to-3D frameworks. For example, we can extract assets from the static phase of TC4D. One example is shown in \cref{fig:text23d}.

\begin{figure}[t]
    \centering
    \includegraphics[width=\linewidth]{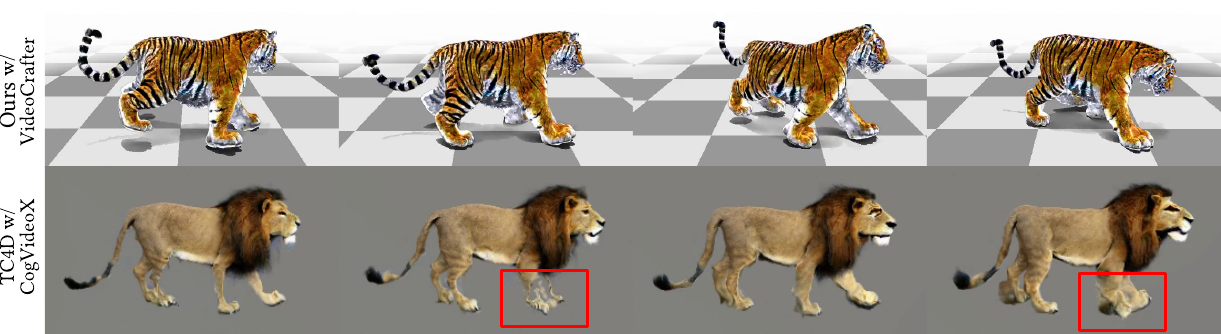}
    \caption{Ablation study on the base video diffusion model.}
    \label{fig:video_model}
\end{figure}

\vspace{-1em}
\paragraph{Video Diffusion Model.}
The video model and the articulated representation both improve the quality of synthesized motions. In \cref{fig:video_model}, our method with VideoCrafter generates alternating leg motions but suffers from severe foot-skating issues. TC4D with CogVideoX still fails to produce alternating leg motions and exhibits blurry artifacts.

\section{Conclusion}

In this paper, we present AKD, a text-driven method for articulated motion synthesis powered by large video diffusion models. By distilling the motion knowledge from a video generation model, our model offers an alternative to laborious motion authoring in traditional animation pipelines.
The low-DoF, skeleton-based parameterization of motion allows the distillation process to focus more on overall articulated motion patterns rather than local-scale shape deformations, resolving the common issue of shape inconsistency as in previous work on 4D generation, and thus results in improved physical plausibility of the generated motion. 
We demonstrate that the generated skeleton motions can be transferred into simulation environments via physics-based motion tracking. We hope this work can inspire future research in generating labeled data for robotics applications. 

Our approach has several limitations that point to future work. The visual quality of our generated assets is still suboptimal; improving the visual quality can help reduce the gap between rendered outputs and the realistic video distributions encoded in video models. The diversity of motions produced by our method largely depends on the video model’s ability to synthesize desired motions; improved video priors in the future could enhance this diversity. Our method focuses on the articulated motion of objects, and may not be suitable for other types of deformation such as soft-body dynamics. Additionally, our pipeline assumes a manually rigged skeleton. To scale up our method, future work can leverage automatic rigging techniques such as RigNet \cite{xu2020rignet} to generalize to diverse character types, and can use more efficient sampling techniques for video models to accelerate convergence.

{
    \small
    \bibliographystyle{ieeenat_fullname}
    \bibliography{main}

\begin{thebibliography}{67}
\providecommand{\natexlab}[1]{#1}
\providecommand{\url}[1]{\texttt{#1}}
\expandafter\ifx\csname urlstyle\endcsname\relax
  \providecommand{\doi}[1]{doi: #1}\else
  \providecommand{\doi}{doi: \begingroup \urlstyle{rm}\Url}\fi

\bibitem[Bahmani et~al.(2024)Bahmani, Skorokhodov, Rong, Wetzstein, Guibas, Wonka, Tulyakov, Park, Tagliasacchi, and Lindell]{bahmani20244d}
Sherwin Bahmani, Ivan Skorokhodov, Victor Rong, Gordon Wetzstein, Leonidas Guibas, Peter Wonka, Sergey Tulyakov, Jeong~Joon Park, Andrea Tagliasacchi, and David~B Lindell.
\newblock 4d-fy: Text-to-4d generation using hybrid score distillation sampling.
\newblock In \emph{Proceedings of the IEEE/CVF Conference on Computer Vision and Pattern Recognition}, pages 7996--8006, 2024.

\bibitem[Bahmani et~al.(2025)Bahmani, Liu, Yifan, Skorokhodov, Rong, Liu, Liu, Park, Tulyakov, Wetzstein, et~al.]{bahmani2025tc4d}
Sherwin Bahmani, Xian Liu, Wang Yifan, Ivan Skorokhodov, Victor Rong, Ziwei Liu, Xihui Liu, Jeong~Joon Park, Sergey Tulyakov, Gordon Wetzstein, et~al.
\newblock Tc4d: Trajectory-conditioned text-to-4d generation.
\newblock In \emph{European Conference on Computer Vision}, pages 53--72. Springer, 2025.

\bibitem[Bansal et~al.(2024)Bansal, Lin, Xie, Zong, Yarom, Bitton, Jiang, Sun, Chang, and Grover]{bansal2024videophy}
Hritik Bansal, Zongyu Lin, Tianyi Xie, Zeshun Zong, Michal Yarom, Yonatan Bitton, Chenfanfu Jiang, Yizhou Sun, Kai-Wei Chang, and Aditya Grover.
\newblock Videophy: Evaluating physical commonsense for video generation.
\newblock \emph{arXiv preprint arXiv:2406.03520}, 2024.

\bibitem[Baran and Popovi{\'c}(2007)]{baran2007automatic}
Ilya Baran and Jovan Popovi{\'c}.
\newblock Automatic rigging and animation of 3d characters.
\newblock \emph{ACM Transactions on graphics (TOG)}, 26\penalty0 (3):\penalty0 72--es, 2007.

\bibitem[{Blender Online Community}(2024)]{blender}
{Blender Online Community}.
\newblock \emph{Blender - a 3D modelling and rendering package}.
\newblock Blender Foundation, 2024.

\bibitem[Chen et~al.(2024)Chen, Zhang, Cun, Xia, Wang, Weng, and Shan]{chen2024videocrafter2}
Haoxin Chen, Yong Zhang, Xiaodong Cun, Menghan Xia, Xintao Wang, Chao Weng, and Ying Shan.
\newblock Videocrafter2: Overcoming data limitations for high-quality video diffusion models.
\newblock In \emph{Proceedings of the IEEE/CVF Conference on Computer Vision and Pattern Recognition}, pages 7310--7320, 2024.

\bibitem[Chen et~al.(2016)Chen, Xu, Zhang, and Guestrin]{chen2016training}
Tianqi Chen, Bing Xu, Chiyuan Zhang, and Carlos Guestrin.
\newblock Training deep nets with sublinear memory cost.
\newblock \emph{arXiv preprint arXiv:1604.06174}, 2016.

\bibitem[Denavit and Hartenberg(1955)]{denavit1955kinematic}
Jacques Denavit and Richard~S Hartenberg.
\newblock A kinematic notation for lower-pair mechanisms based on matrices.
\newblock 1955.

\bibitem[Feng et~al.(2024)Feng, Shang, Li, Shao, Jiang, and Yang]{feng2024pie}
Yutao Feng, Yintong Shang, Xuan Li, Tianjia Shao, Chenfanfu Jiang, and Yin Yang.
\newblock Pie-nerf: Physics-based interactive elastodynamics with nerf.
\newblock In \emph{Proceedings of the IEEE/CVF Conference on Computer Vision and Pattern Recognition}, pages 4450--4461, 2024.

\bibitem[Giebenhain et~al.(2024)Giebenhain, Kirschstein, R{\"u}nz, Agapito, and Nie{\ss}ner]{giebenhain2024npga}
Simon Giebenhain, Tobias Kirschstein, Martin R{\"u}nz, Lourdes Agapito, and Matthias Nie{\ss}ner.
\newblock Npga: Neural parametric gaussian avatars.
\newblock \emph{arXiv preprint arXiv:2405.19331}, 2024.

\bibitem[Goel et~al.(2023)Goel, Pavlakos, Rajasegaran, Kanazawa, and Malik]{goel2023humans}
Shubham Goel, Georgios Pavlakos, Jathushan Rajasegaran, Angjoo Kanazawa, and Jitendra Malik.
\newblock Humans in 4d: Reconstructing and tracking humans with transformers.
\newblock In \emph{Proceedings of the IEEE/CVF International Conference on Computer Vision}, pages 14783--14794, 2023.

\bibitem[Gu et~al.(2024)Gu, Yang, Pan, Zhu, and Zhang]{gu2024tetrahedron}
Chun Gu, Zeyu Yang, Zijie Pan, Xiatian Zhu, and Li Zhang.
\newblock Tetrahedron splatting for 3d generation.
\newblock In \emph{NeurIPS}, 2024.

\bibitem[Guo et~al.(2023)Guo, Liu, Shao, Laforte, Voleti, Luo, Chen, Zou, Wang, Cao, and Zhang]{threestudio2023}
Yuan-Chen Guo, Ying-Tian Liu, Ruizhi Shao, Christian Laforte, Vikram Voleti, Guan Luo, Chia-Hao Chen, Zi-Xin Zou, Chen Wang, Yan-Pei Cao, and Song-Hai Zhang.
\newblock threestudio: A unified framework for 3d content generation.
\newblock \url{https://github.com/threestudio-project/threestudio}, 2023.

\bibitem[Ho et~al.(2020)Ho, Jain, and Abbeel]{ho2020denoising}
Jonathan Ho, Ajay Jain, and Pieter Abbeel.
\newblock Denoising diffusion probabilistic models.
\newblock \emph{Advances in neural information processing systems}, 33:\penalty0 6840--6851, 2020.

\bibitem[Hu et~al.(2024)Hu, Hu, and Liu]{hu2024gauhuman}
Shoukang Hu, Tao Hu, and Ziwei Liu.
\newblock Gauhuman: Articulated gaussian splatting from monocular human videos.
\newblock In \emph{Proceedings of the IEEE/CVF Conference on Computer Vision and Pattern Recognition}, pages 20418--20431, 2024.

\bibitem[Jakab et~al.(2024)Jakab, Li, Wu, Rupprecht, and Vedaldi]{jakab2024farm3d}
Tomas Jakab, Ruining Li, Shangzhe Wu, Christian Rupprecht, and Andrea Vedaldi.
\newblock Farm3d: Learning articulated 3d animals by distilling 2d diffusion.
\newblock In \emph{2024 International Conference on 3D Vision (3DV)}, pages 852--861. IEEE, 2024.

\bibitem[Jiang et~al.(2023{\natexlab{a}})Jiang, Chen, Liu, Yu, Yu, and Chen]{jiang2023motiongpt}
Biao Jiang, Xin Chen, Wen Liu, Jingyi Yu, Gang Yu, and Tao Chen.
\newblock Motiongpt: Human motion as a foreign language.
\newblock \emph{Advances in Neural Information Processing Systems}, 36:\penalty0 20067--20079, 2023{\natexlab{a}}.

\bibitem[Jiang et~al.(2023{\natexlab{b}})Jiang, Zhang, Gao, Hu, and Yao]{jiang2023consistent4d}
Yanqin Jiang, Li Zhang, Jin Gao, Weimin Hu, and Yao Yao.
\newblock Consistent4d: Consistent 360 $\{$$\backslash$deg$\}$ dynamic object generation from monocular video.
\newblock \emph{arXiv preprint arXiv:2311.02848}, 2023{\natexlab{b}}.

\bibitem[Jiang et~al.(2024)Jiang, Yu, Cao, Wang, Hu, and Gao]{jiang2024animate3d}
Yanqin Jiang, Chaohui Yu, Chenjie Cao, Fan Wang, Weiming Hu, and Jin Gao.
\newblock Animate3d: Animating any 3d model with multi-view video diffusion.
\newblock \emph{arXiv preprint arXiv:2407.11398}, 2024.

\bibitem[Kanazawa et~al.(2018)Kanazawa, Black, Jacobs, and Malik]{kanazawa2018end}
Angjoo Kanazawa, Michael~J Black, David~W Jacobs, and Jitendra Malik.
\newblock End-to-end recovery of human shape and pose.
\newblock In \emph{Proceedings of the IEEE conference on computer vision and pattern recognition}, pages 7122--7131, 2018.

\bibitem[Kerbl et~al.(2023)Kerbl, Kopanas, Leimk{\"u}hler, and Drettakis]{kerbl20233d}
Bernhard Kerbl, Georgios Kopanas, Thomas Leimk{\"u}hler, and George Drettakis.
\newblock 3d gaussian splatting for real-time radiance field rendering.
\newblock \emph{ACM Trans. Graph.}, 42\penalty0 (4):\penalty0 139--1, 2023.

\bibitem[Kocabas et~al.(2024)Kocabas, Chang, Gabriel, Tuzel, and Ranjan]{kocabas2024hugs}
Muhammed Kocabas, Jen-Hao~Rick Chang, James Gabriel, Oncel Tuzel, and Anurag Ranjan.
\newblock Hugs: Human gaussian splats.
\newblock In \emph{Proceedings of the IEEE/CVF conference on computer vision and pattern recognition}, pages 505--515, 2024.

\bibitem[Lei et~al.(2024)Lei, Wang, Pavlakos, Liu, and Daniilidis]{lei2024gart}
Jiahui Lei, Yufu Wang, Georgios Pavlakos, Lingjie Liu, and Kostas Daniilidis.
\newblock Gart: Gaussian articulated template models.
\newblock In \emph{Proceedings of the IEEE/CVF Conference on Computer Vision and Pattern Recognition}, pages 19876--19887, 2024.

\bibitem[Li et~al.(2024)Li, Litvak, Li, Zhang, Jakab, Rupprecht, Wu, Vedaldi, and Wu]{li2024learning}
Zizhang Li, Dor Litvak, Ruining Li, Yunzhi Zhang, Tomas Jakab, Christian Rupprecht, Shangzhe Wu, Andrea Vedaldi, and Jiajun Wu.
\newblock Learning the 3d fauna of the web.
\newblock In \emph{Proceedings of the IEEE/CVF Conference on Computer Vision and Pattern Recognition}, pages 9752--9762, 2024.

\bibitem[Lin et~al.(2014)Lin, Maire, Belongie, Hays, Perona, Ramanan, Doll{\'a}r, and Zitnick]{lin2014microsoft}
Tsung-Yi Lin, Michael Maire, Serge Belongie, James Hays, Pietro Perona, Deva Ramanan, Piotr Doll{\'a}r, and C~Lawrence Zitnick.
\newblock Microsoft coco: Common objects in context.
\newblock In \emph{Computer Vision--ECCV 2014: 13th European Conference, Zurich, Switzerland, September 6-12, 2014, Proceedings, Part V 13}, pages 740--755. Springer, 2014.

\bibitem[Ling et~al.(2024)Ling, Kim, Torralba, Fidler, and Kreis]{ling2024align}
Huan Ling, Seung~Wook Kim, Antonio Torralba, Sanja Fidler, and Karsten Kreis.
\newblock Align your gaussians: Text-to-4d with dynamic 3d gaussians and composed diffusion models.
\newblock In \emph{Proceedings of the IEEE/CVF Conference on Computer Vision and Pattern Recognition}, pages 8576--8588, 2024.

\bibitem[Liu et~al.(2024)Liu, Wu, Liu, Liu, Zhao, Feng, Ding, and Wang]{liu2024gva}
Xinqi Liu, Chenming Wu, Jialun Liu, Xing Liu, Chen Zhao, Haocheng Feng, Errui Ding, and Jingdong Wang.
\newblock Gva: Reconstructing vivid 3d gaussian avatars from monocular videos.
\newblock \emph{CoRR}, 2024.

\bibitem[Loper et~al.(2023)Loper, Mahmood, Romero, Pons-Moll, and Black]{loper2023smpl}
Matthew Loper, Naureen Mahmood, Javier Romero, Gerard Pons-Moll, and Michael~J Black.
\newblock Smpl: A skinned multi-person linear model.
\newblock In \emph{Seminal Graphics Papers: Pushing the Boundaries, Volume 2}, pages 851--866. 2023.

\bibitem[Macklin(2022)]{warp2022}
Miles Macklin.
\newblock Warp: A high-performance python framework for gpu simulation and graphics.
\newblock \url{https://github.com/nvidia/warp}, 2022.
\newblock NVIDIA GPU Technology Conference (GTC).

\bibitem[Magnenat et~al.(1988)Magnenat, Laperri{\`e}re, and Thalmann]{magnenat1988joint}
Thalmann Magnenat, Richard Laperri{\`e}re, and Daniel Thalmann.
\newblock Joint-dependent local deformations for hand animation and object grasping.
\newblock In \emph{Proceedings of Graphics Interface'88}, pages 26--33. Canadian Inf. Process. Soc, 1988.

\bibitem[Mahmood et~al.(2019)Mahmood, Ghorbani, Troje, Pons-Moll, and Black]{mahmood2019amass}
Naureen Mahmood, Nima Ghorbani, Nikolaus~F Troje, Gerard Pons-Moll, and Michael~J Black.
\newblock Amass: Archive of motion capture as surface shapes.
\newblock In \emph{Proceedings of the IEEE/CVF international conference on computer vision}, pages 5442--5451, 2019.

\bibitem[Mildenhall et~al.(2021)Mildenhall, Srinivasan, Tancik, Barron, Ramamoorthi, and Ng]{mildenhall2021nerf}
Ben Mildenhall, Pratul~P Srinivasan, Matthew Tancik, Jonathan~T Barron, Ravi Ramamoorthi, and Ren Ng.
\newblock Nerf: Representing scenes as neural radiance fields for view synthesis.
\newblock \emph{Communications of the ACM}, 65\penalty0 (1):\penalty0 99--106, 2021.

\bibitem[Moreau et~al.(2024)Moreau, Song, Dhamo, Shaw, Zhou, and P{\'e}rez-Pellitero]{moreau2024human}
Arthur Moreau, Jifei Song, Helisa Dhamo, Richard Shaw, Yiren Zhou, and Eduardo P{\'e}rez-Pellitero.
\newblock Human gaussian splatting: Real-time rendering of animatable avatars.
\newblock In \emph{Proceedings of the IEEE/CVF Conference on Computer Vision and Pattern Recognition}, pages 788--798, 2024.

\bibitem[M{\"u}ller et~al.(2022)M{\"u}ller, Evans, Schied, and Keller]{muller2022instant}
Thomas M{\"u}ller, Alex Evans, Christoph Schied, and Alexander Keller.
\newblock Instant neural graphics primitives with a multiresolution hash encoding.
\newblock \emph{ACM transactions on graphics (TOG)}, 41\penalty0 (4):\penalty0 1--15, 2022.

\bibitem[Park et~al.(2021)Park, Sinha, Barron, Bouaziz, Goldman, Seitz, and Martin-Brualla]{park2021nerfies}
Keunhong Park, Utkarsh Sinha, Jonathan~T Barron, Sofien Bouaziz, Dan~B Goldman, Steven~M Seitz, and Ricardo Martin-Brualla.
\newblock Nerfies: Deformable neural radiance fields.
\newblock In \emph{Proceedings of the IEEE/CVF International Conference on Computer Vision}, pages 5865--5874, 2021.

\bibitem[Poole et~al.(2023)Poole, Jain, Barron, and Mildenhall]{poole2023dreamfusion}
Ben Poole, Ajay Jain, Jonathan~T. Barron, and Ben Mildenhall.
\newblock Dreamfusion: Text-to-3d using 2d diffusion.
\newblock In \emph{The Eleventh International Conference on Learning Representations}, 2023.

\bibitem[Pumarola et~al.(2021)Pumarola, Corona, Pons-Moll, and Moreno-Noguer]{pumarola2021d}
Albert Pumarola, Enric Corona, Gerard Pons-Moll, and Francesc Moreno-Noguer.
\newblock D-nerf: Neural radiance fields for dynamic scenes.
\newblock In \emph{Proceedings of the IEEE/CVF Conference on Computer Vision and Pattern Recognition}, pages 10318--10327, 2021.

\bibitem[Qiao et~al.(2023)Qiao, Gao, Xu, Feng, Huang, and Lin]{qiao2023dynamic}
Yi-Ling Qiao, Alexander Gao, Yiran Xu, Yue Feng, Jia-Bin Huang, and Ming~C Lin.
\newblock Dynamic mesh-aware radiance fields.
\newblock In \emph{Proceedings of the IEEE/CVF International Conference on Computer Vision}, pages 385--396, 2023.

\bibitem[Raab et~al.(2024)Raab, Leibovitch, Tevet, Arar, Bermano, and Cohen-Or]{raab2024single}
Sigal Raab, Inbal Leibovitch, Guy Tevet, Moab Arar, Amit~Haim Bermano, and Daniel Cohen-Or.
\newblock Single motion diffusion.
\newblock In \emph{ICLR}, 2024.

\bibitem[Ren et~al.(2023)Ren, Pan, Tang, Zhang, Cao, Zeng, and Liu]{ren2023dreamgaussian4d}
Jiawei Ren, Liang Pan, Jiaxiang Tang, Chi Zhang, Ang Cao, Gang Zeng, and Ziwei Liu.
\newblock Dreamgaussian4d: Generative 4d gaussian splatting.
\newblock \emph{arXiv preprint arXiv:2312.17142}, 2023.

\bibitem[Ren et~al.(2024)Ren, Xie, Mirzaei, Liang, Zeng, Kreis, Liu, Torralba, Fidler, Kim, et~al.]{ren2024l4gm}
Jiawei Ren, Kevin Xie, Ashkan Mirzaei, Hanxue Liang, Xiaohui Zeng, Karsten Kreis, Ziwei Liu, Antonio Torralba, Sanja Fidler, Seung~Wook Kim, et~al.
\newblock L4gm: Large 4d gaussian reconstruction model.
\newblock \emph{arXiv preprint arXiv:2406.10324}, 2024.

\bibitem[Salimans and Ho(2022)]{salimans2022progressive}
Tim Salimans and Jonathan Ho.
\newblock Progressive distillation for fast sampling of diffusion models.
\newblock In \emph{International Conference on Learning Representations}, 2022.

\bibitem[Shao et~al.(2024)Shao, Wang, Li, Wang, Lin, Zhang, Fan, and Wang]{shao2024splattingavatar}
Zhijing Shao, Zhaolong Wang, Zhuang Li, Duotun Wang, Xiangru Lin, Yu Zhang, Mingming Fan, and Zeyu Wang.
\newblock Splattingavatar: Realistic real-time human avatars with mesh-embedded gaussian splatting.
\newblock In \emph{Proceedings of the IEEE/CVF Conference on Computer Vision and Pattern Recognition}, pages 1606--1616, 2024.

\bibitem[Singer et~al.(2023)Singer, Sheynin, Polyak, Ashual, Makarov, Kokkinos, Goyal, Vedaldi, Parikh, Johnson, et~al.]{singer2023text}
Uriel Singer, Shelly Sheynin, Adam Polyak, Oron Ashual, Iurii Makarov, Filippos Kokkinos, Naman Goyal, Andrea Vedaldi, Devi Parikh, Justin Johnson, et~al.
\newblock Text-to-4d dynamic scene generation.
\newblock \emph{arXiv preprint arXiv:2301.11280}, 2023.

\bibitem[Sun et~al.(2025)Sun, Litvak, Zhang, Li, Wu, and Wu]{sun2025ponymation}
Keqiang Sun, Dor Litvak, Yunzhi Zhang, Hongsheng Li, Jiajun Wu, and Shangzhe Wu.
\newblock Ponymation: Learning articulated 3d animal motions from unlabeled online videos.
\newblock In \emph{European Conference on Computer Vision}, pages 100--119. Springer, 2025.

\bibitem[Tan et~al.(2024)Tan, Xiang, Tulsiani, Ramanan, and Yang]{tan2024dressrecon}
Jeff Tan, Donglai Xiang, Shubham Tulsiani, Deva Ramanan, and Gengshan Yang.
\newblock Dressrecon: Freeform 4d human reconstruction from monocular video.
\newblock \emph{arXiv preprint arXiv:2409.20563}, 2024.

\bibitem[Tevet et~al.(2023)Tevet, Raab, Gordon, Shafir, Cohen-or, and Bermano]{tevet2023human}
Guy Tevet, Sigal Raab, Brian Gordon, Yoni Shafir, Daniel Cohen-or, and Amit~Haim Bermano.
\newblock Human motion diffusion model.
\newblock In \emph{The Eleventh International Conference on Learning Representations}, 2023.

\bibitem[Uzolas et~al.(2024)Uzolas, Eisemann, and Kellnhofer]{uzolas2024motiondreamer}
Lukas Uzolas, Elmar Eisemann, and Petr Kellnhofer.
\newblock Motiondreamer: Exploring semantic video diffusion features for zero-shot 3d mesh animation, 2024.

\bibitem[Wu et~al.(2023)Wu, Li, Jakab, Rupprecht, and Vedaldi]{wu2023magicpony}
Shangzhe Wu, Ruining Li, Tomas Jakab, Christian Rupprecht, and Andrea Vedaldi.
\newblock Magicpony: Learning articulated 3d animals in the wild.
\newblock In \emph{Proceedings of the IEEE/CVF Conference on Computer Vision and Pattern Recognition}, pages 8792--8802, 2023.

\bibitem[Wu et~al.(2025)Wu, Yu, Jiang, Cao, Wang, and Bai]{wu2025sc4d}
Zijie Wu, Chaohui Yu, Yanqin Jiang, Chenjie Cao, Fan Wang, and Xiang Bai.
\newblock Sc4d: Sparse-controlled video-to-4d generation and motion transfer.
\newblock In \emph{European Conference on Computer Vision}, pages 361--379. Springer, 2025.

\bibitem[Xiang et~al.(2019)Xiang, Joo, and Sheikh]{xiang2019monocular}
Donglai Xiang, Hanbyul Joo, and Yaser Sheikh.
\newblock Monocular total capture: Posing face, body, and hands in the wild.
\newblock In \emph{Proceedings of the IEEE/CVF conference on computer vision and pattern recognition}, pages 10965--10974, 2019.

\bibitem[Xie et~al.(2024{\natexlab{a}})Xie, Zong, Qiu, Li, Feng, Yang, and Jiang]{xie2024physgaussian}
Tianyi Xie, Zeshun Zong, Yuxing Qiu, Xuan Li, Yutao Feng, Yin Yang, and Chenfanfu Jiang.
\newblock Physgaussian: Physics-integrated 3d gaussians for generative dynamics.
\newblock In \emph{Proceedings of the IEEE/CVF Conference on Computer Vision and Pattern Recognition}, pages 4389--4398, 2024{\natexlab{a}}.

\bibitem[Xie et~al.(2024{\natexlab{b}})Xie, Yao, Voleti, Jiang, and Jampani]{xie2024sv4d}
Yiming Xie, Chun-Han Yao, Vikram Voleti, Huaizu Jiang, and Varun Jampani.
\newblock Sv4d: Dynamic 3d content generation with multi-frame and multi-view consistency.
\newblock \emph{arXiv preprint arXiv:2407.17470}, 2024{\natexlab{b}}.

\bibitem[Xu et~al.(2024)Xu, Chen, Li, Zhang, Wang, Zheng, and Liu]{xu2024gaussian}
Yuelang Xu, Benwang Chen, Zhe Li, Hongwen Zhang, Lizhen Wang, Zerong Zheng, and Yebin Liu.
\newblock Gaussian head avatar: Ultra high-fidelity head avatar via dynamic gaussians.
\newblock In \emph{Proceedings of the IEEE/CVF Conference on Computer Vision and Pattern Recognition}, pages 1931--1941, 2024.

\bibitem[Xu et~al.(2020)Xu, Zhou, Kalogerakis, Landreth, and Singh]{xu2020rignet}
Zhan Xu, Yang Zhou, Evangelos Kalogerakis, Chris Landreth, and Karan Singh.
\newblock Rignet: Neural rigging for articulated characters.
\newblock \emph{arXiv preprint arXiv:2005.00559}, 2020.

\bibitem[Yang et~al.(2022)Yang, Vo, Neverova, Ramanan, Vedaldi, and Joo]{yang2022banmo}
Gengshan Yang, Minh Vo, Natalia Neverova, Deva Ramanan, Andrea Vedaldi, and Hanbyul Joo.
\newblock Banmo: Building animatable 3d neural models from many casual videos.
\newblock In \emph{Proceedings of the IEEE/CVF Conference on Computer Vision and Pattern Recognition}, pages 2863--2873, 2022.

\bibitem[Yang et~al.(2023{\natexlab{a}})Yang, Wang, Reddy, and Ramanan]{yang2023reconstructing}
Gengshan Yang, Chaoyang Wang, N~Dinesh Reddy, and Deva Ramanan.
\newblock Reconstructing animatable categories from videos.
\newblock In \emph{Proceedings of the IEEE/CVF Conference on Computer Vision and Pattern Recognition}, pages 16995--17005, 2023{\natexlab{a}}.

\bibitem[Yang et~al.(2023{\natexlab{b}})Yang, Yang, Zhang, Manchester, and Ramanan]{yang2023ppr}
Gengshan Yang, Shuo Yang, John~Z Zhang, Zachary Manchester, and Deva Ramanan.
\newblock Ppr: Physically plausible reconstruction from monocular videos.
\newblock In \emph{Proceedings of the IEEE/CVF International Conference on Computer Vision}, pages 3914--3924, 2023{\natexlab{b}}.

\bibitem[Yang et~al.(2024)Yang, Teng, Zheng, Ding, Huang, Xu, Yang, Hong, Zhang, Feng, et~al.]{yang2024cogvideox}
Zhuoyi Yang, Jiayan Teng, Wendi Zheng, Ming Ding, Shiyu Huang, Jiazheng Xu, Yuanming Yang, Wenyi Hong, Xiaohan Zhang, Guanyu Feng, et~al.
\newblock Cogvideox: Text-to-video diffusion models with an expert transformer.
\newblock \emph{arXiv preprint arXiv:2408.06072}, 2024.

\bibitem[Yao et~al.(2022)Yao, Hung, Li, Rubinstein, Yang, and Jampani]{yao2022lassie}
Chun-Han Yao, Wei-Chih Hung, Yuanzhen Li, Michael Rubinstein, Ming-Hsuan Yang, and Varun Jampani.
\newblock Lassie: Learning articulated shapes from sparse image ensemble via 3d part discovery.
\newblock \emph{Advances in Neural Information Processing Systems}, 35:\penalty0 15296--15308, 2022.

\bibitem[Yao et~al.(2023)Yao, Hung, Li, Rubinstein, Yang, and Jampani]{yao2023hi}
Chun-Han Yao, Wei-Chih Hung, Yuanzhen Li, Michael Rubinstein, Ming-Hsuan Yang, and Varun Jampani.
\newblock Hi-lassie: High-fidelity articulated shape and skeleton discovery from sparse image ensemble.
\newblock In \emph{Proceedings of the IEEE/CVF Conference on Computer Vision and Pattern Recognition}, pages 4853--4862, 2023.

\bibitem[Yu et~al.(2024)Yu, Chen, Huang, Sattler, and Geiger]{yu2024mip}
Zehao Yu, Anpei Chen, Binbin Huang, Torsten Sattler, and Andreas Geiger.
\newblock Mip-splatting: Alias-free 3d gaussian splatting.
\newblock In \emph{Proceedings of the IEEE/CVF Conference on Computer Vision and Pattern Recognition}, pages 19447--19456, 2024.

\bibitem[Yuan et~al.(2023)Yuan, Song, Iqbal, Vahdat, and Kautz]{yuan2023physdiff}
Ye Yuan, Jiaming Song, Umar Iqbal, Arash Vahdat, and Jan Kautz.
\newblock Physdiff: Physics-guided human motion diffusion model.
\newblock In \emph{Proceedings of the IEEE/CVF international conference on computer vision}, pages 16010--16021, 2023.

\bibitem[Zeng et~al.(2025)Zeng, Jiang, Zhu, Lu, Lin, Zhu, Hu, Cao, and Yao]{zeng2025stag4d}
Yifei Zeng, Yanqin Jiang, Siyu Zhu, Yuanxun Lu, Youtian Lin, Hao Zhu, Weiming Hu, Xun Cao, and Yao Yao.
\newblock Stag4d: Spatial-temporal anchored generative 4d gaussians.
\newblock In \emph{European Conference on Computer Vision}, pages 163--179. Springer, 2025.

\bibitem[Zhang et~al.(2025)Zhang, Yu, Wu, Feng, Zheng, Snavely, Wu, and Freeman]{zhang2025physdreamer}
Tianyuan Zhang, Hong-Xing Yu, Rundi Wu, Brandon~Y Feng, Changxi Zheng, Noah Snavely, Jiajun Wu, and William~T Freeman.
\newblock Physdreamer: Physics-based interaction with 3d objects via video generation.
\newblock In \emph{European Conference on Computer Vision}, pages 388--406. Springer, 2025.

\bibitem[Zheng et~al.(2024{\natexlab{a}})Zheng, Li, Nagano, Liu, Hilliges, and De~Mello]{zheng2024unified}
Yufeng Zheng, Xueting Li, Koki Nagano, Sifei Liu, Otmar Hilliges, and Shalini De~Mello.
\newblock A unified approach for text-and image-guided 4d scene generation.
\newblock In \emph{Proceedings of the IEEE/CVF Conference on Computer Vision and Pattern Recognition}, pages 7300--7309, 2024{\natexlab{a}}.

\bibitem[Zheng et~al.(2024{\natexlab{b}})Zheng, Zhao, Yang, Yifan, Xiang, Dubost, Lagun, Beeler, Tombari, Guibas, et~al.]{zheng2024physavatar}
Yang Zheng, Qingqing Zhao, Guandao Yang, Wang Yifan, Donglai Xiang, Florian Dubost, Dmitry Lagun, Thabo Beeler, Federico Tombari, Leonidas Guibas, et~al.
\newblock Physavatar: Learning the physics of dressed 3d avatars from visual observations.
\newblock \emph{arXiv preprint arXiv:2404.04421}, 2024{\natexlab{b}}.

\end{thebibliography}
}

\appendix

\section*{Appendix}

\begin{figure*}[t!]
    \centering
    \includegraphics[width=\linewidth]{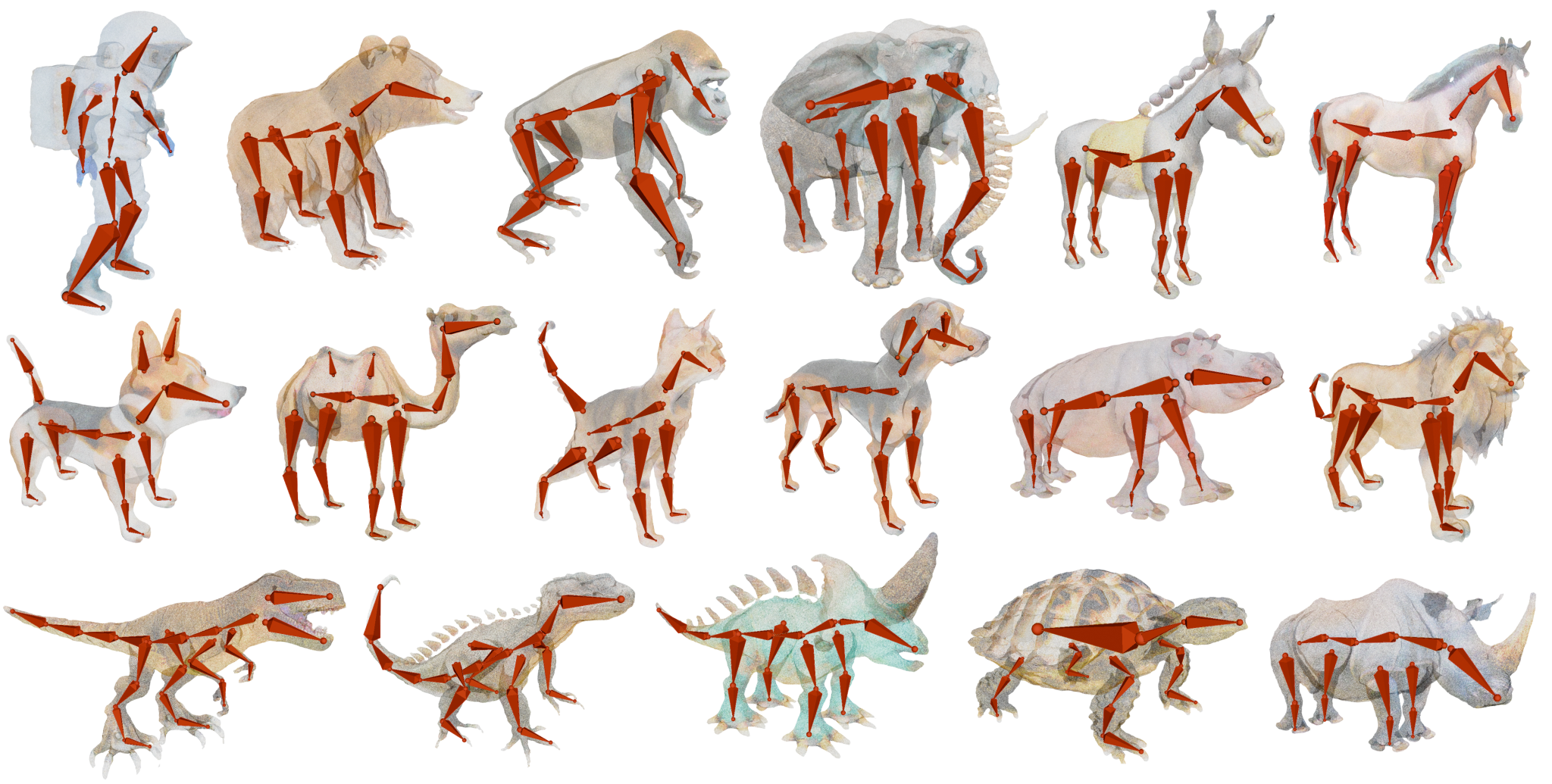}
    \caption{Gallery of skeleton systems from our experiments.}
    \label{fig:skeleton}
\end{figure*}

\section{Automatic Skinning Weight Computation}
Assigning skinning weights to Gaussian kernels is not a straightforward task. For example, a kernel close to a bone actually should not be affected by it if the shortest segment from the kernel passes the outside region of the object. As a concrete example, Gaussian kernels on one foot of a human should not be influenced by the other foot. While using learnable weights could address this \cite{moreau2024human, kocabas2024hugs, hu2024gauhuman}, it would undermine the degree-of-freedom (DoF) reduction achieved by the rigging system. To resolve ambiguities in weight assignment, we use a reference mesh that aligns with the geometry of 3DGS to define weights on the mesh surface, and then transfer these weights to the Gaussian kernels. This mesh is typically an output form from text-to-3D frameworks or can be generated using mesh extraction tools. 
For automatic weight computation on the mesh, we use the widely adopted auto-rigging system Pinocchio \cite{baran2007automatic}. Pinocchio conceptualizes weight computation on a simply connected mesh as a heat diffusion process along the mesh surface, accounting for bone visibilities blocked by other surface parts. Specifically, for the bone $b$, the weight contribution vector $\bm w^b$ on vertices is computed by solving the following Poisson equation:
\begin{equation} \label{eq:Pinocchio}
    - \Delta \bm{w}^b + \mathbf{H}^b \bm{w}^b = \mathbf{H}^b \mathbf{p}^b.
\end{equation}
Here, $\Delta $ is the cotangent surface Laplacian operator. $\mathbf{p}^b_j = 1$ only if the closest bone to the vertex $j$ is $b$. $\bm H^b$ is a diagonal matrix with entries $\bm H_{jj}^b = c/d^2_j$ only if bone $b$ is visible from vertex $j$ within the mesh, where $c$ is a user-defined constant and $d_j$ is the distance from vertex $j$ to bone $b$. The visibility is determined by whether the segment from the vertex to its closest point on the bone is completely enclosed within the mesh. This equation can be interpreted as follows: the bone first transfers heat to its visible vertex, and then the heat diffuses along the surface. The resulting heat distribution represents the weight field for that bone. We extend Pinocchio to support noisy mesh inputs that may have multiple components by manually setting the visibility of a connected component to its nearest bone as true if all bones are invisible from that component. This can prevent the system matrix of \cref{eq:Pinocchio} from being singular and allows the outlier components to be controlled by their nearest bones.

By concatenating the weight contributions of each bone, we obtain the weight matrix $\bm W \in \mathbb{R}^{V \times B}$, where $V$ is the number of vertices. Each row of $\bm W$ represents the skinning weight vector at a given vertex. These weight vectors can be interpolated to arbitrary surface points on the mesh using barycentric interpolation. Since Gaussian kernels generated during reconstruction are typically located near the geometric surface, we identify weight vectors of Gaussian kernels with their nearest surface points on the mesh.

\section{SDS Gradient for V-Prediction Diffusion Models}
The video diffusion model in our pipeline, CogVideoX-5B \cite{yang2024cogvideox}, is a $v$-prediction diffusion model \cite{salimans2022progressive}, where the model predicts the so-called velocity instead of noise. The diffusion loss for training is 
\begin{equation}
\small
    \mathcal{L}_{\text{Diff}}(\theta, \bm z,  y) = \mathbb{E}_{t, \bm\epsilon} \left[ \frac{1}{1-\alpha_t} \left\| \bm z - \hat{\bm z} \right\|_2^2 \right],
\end{equation}
where $\bm z$ is the latent code of a video, $\hat{\bm z} = \sqrt{\alpha_t} \bm z_t - \sqrt{1 - \alpha_t} \bm v_{\theta}(\bm z_t; t, y)$, $\bm z_t =\sqrt{\alpha_t} \bm z + \sqrt{1-\alpha_t} \bm \epsilon $, and $\bm v_{\theta}$ is a large transformer. Taking the derivative of $\mathcal{L}_{\text{Diff}}$ w.r.t. $\bm z$, we get
\begin{equation}
\footnotesize
\begin{split}
    \nabla_{\bm z} \mathcal{L}_{\text{Diff}} =& \mathbb{E}_{t, \bm\epsilon} \left[ \frac{1}{1-\alpha_t} (I - \frac{\partial \hat{\bm z}}{\partial \bm z}) \left( \bm z - \hat{\bm z} \right) \right]\\
    =&\mathbb{E}_{t, \bm\epsilon} \left[ \frac{1}{1-\alpha_t} (I - \frac{\partial \hat{\bm z}}{\partial \bm z_t}\frac{\partial {\bm z_t}}{\partial \bm z} - \frac{\partial \hat{\bm z}}{\partial \bm v_\theta}\frac{\partial \bm v_\theta}{\partial \bm z}) \left( \bm z - \hat{\bm z} \right) \right].
\end{split}
\end{equation}
Here we omit the constant $2$ that arise from the derivative of the square function for notation simplicity.
Following the SDS gradients for U-Net-based diffusions \cite{poole2023dreamfusion}, where terms involving the gradients of U-Nets are omitted, we similarly omit $\frac{\partial \hat{\bm z}}{\partial \bm v_\theta}\frac{\partial \bm v_\theta}{\partial \bm z}$:
\begin{equation}
\begin{split}
    \nabla_{\bm z} \mathcal{L}_{\text{SDS}}=&\mathbb{E}_{t, \bm\epsilon} \left[ \frac{1}{1-\alpha_t} (1 - (\sqrt{\alpha_t})^2) \left( \bm z - \hat{\bm z} \right) \right] \\
    =&\mathbb{E}_{t, \bm\epsilon} \left[\bm z - \hat{\bm z} \right].
\end{split}
\end{equation}

\section{Shadow Casting.} We can also cast shadows on the ground layer mentioned above to further indicate the spatial relationship between the object and the ground.
When a rendering ray intersects with the ground at a point $\bm P$, the ground color is weighted by this heuristic shadow intensity: $s(\bm P) = 1 - s_{\text{max}} \operatorname{exp}(-\beta d(\bm P))$,
where $d(\bm P)$ is the vertical height from the ray-ground intersection point $\bm P$ to the deformed asset, $s_{\text{max}}$ is the maximum level of shadowing to apply, and $\beta$ is the decay coefficient as the object height above increases. This shadowing approximates a distant, parallel light source positioned vertically above the ground, complemented by diffusive ambient lighting. Incorporating shadows in SDS optimization does not yield substantial improvements in motion synthesis quality, but it can increase the immersiveness when humans evaluate the generated videos. We leave a more thorough investigation of more advanced rendering techniques such as true global illumination to future work.

\section{Implementation Details}
\paragraph{Motion Synthesis} 
The video SDS loss is evaluated with $\text{CFG}=100$. The diffusion time $t$ is sampled uniformly from $[t_{\text{start}}, t_{\text{end}}]$, where $t_{\text{start}} = 0.02$ and  $t_{\text{end}}$ decrease linearly from $0.98$ to $0.5$ over the first 5000 optimization iterations. In the training loss, we set $\lambda_1 = 2\times 10^5$ and $\lambda_2 = 10^7$. A total of 10,000 optimization iterations are performed.

\paragraph{Motion Tracking}
We use Warp~\cite{warp2022} to simulate skeletons as articulated rigid bodies while optimizing the tracking loss and applying gradient clipping with PyTorch. A chunk of simulation substeps is wrapped as a differentiable Pytorch layer using \texttt{torch.autograd.Function}. During the forward pass, data from the PyTorch scope is transferred to Warp, and a Warp gradient tape is initialized and stored to capture the computational graph within the Warp scope during forward simulation. In the backward pass, gradients received from the PyTorch scope are transferred to the Warp scope, and the gradient tape backpropagates the gradients in reverse time. Once the required gradients for the layer's inputs are computed, gradient clipping is applied to them before transferring them back to the PyTorch scope. To ensure that assets remain standing on their own, we introduce a virtual penalty force to keep the root bones of their articulation trees aligned with their initial upward directions. During training, we set $\lambda_3 = 0.2$ and perform 200 optimization iterations.

\section{Skeleton Gallery}
Here, we present a gallery of skeleton systems utilized in our experiments, as shown in \cref{fig:skeleton}. All skeletons are manually crafted in Blender. Mesh representations of assets are initially imported into Blender, followed by the embedding of bones based on their biokinematic structure.

\begin{figure}[t!]
    \centering
    \includegraphics[width=\linewidth]{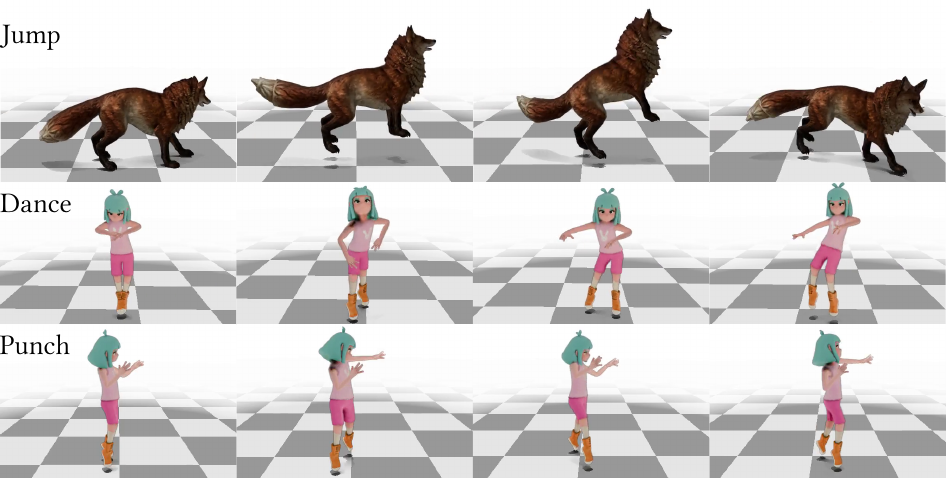}
     \vspace{-1.5em}
    \caption{Additional Motion Variety Experiments}
    \label{fig:motion_diversity}
\end{figure}

\section{Additional Motion Diversity Experiments.}
We find that whether our method can produce diverse motions largely depends on the ability of the video model to synthesize desired full-body motions. In  \cref{fig:motion_diversity}, we show a fox jumping, and a girl dancing and punching. On the other hand, it is difficult for our method to generate animals sitting since CogVideoX struggles with transitions from standing to sitting, as well as precise human motion control such as cartwheeling and raising hands. Additionally, fine-grained motions, such as hand pose variations, are difficult to capture. We found that additional consistent textural descriptions in prompts are crucial for human motion synthesis. We will address these limitations in the revision.

\end{document}